\documentclass[twocolumn]{aastex62}
\usepackage{amsmath}
\usepackage{gensymb}

\usepackage{color}

\received{2019 September 6}
\revised{2019 November 11}
\accepted{2019 November 29}

\shorttitle{Observations and modeling of a confined jet}
\shortauthors{Doyle et al.}

\def\arcsec{\hbox{$^{\prime\prime}$}}

\begin{document}

\title{Observations and 3D MHD Modeling of a Confined Helical Jet Launched by a Filament Eruption}

\correspondingauthor{Lauren Doyle}
\email{lauren.doyle@armagh.ac.uk}

\author{Lauren Doyle}
\affil{Armagh Observatory and Planetarium, College Hill, Armagh, BT61 9DG}
\affiliation{Department of Mathematics, Physics and Electrical Engineering, Northumbria University, Newcastle upon Tyne, NE1 8ST}

\author{Peter F. Wyper}
\affiliation{Department of Mathematical Sciences, Durham University, Durham, DH1 3LE}

\author{Eamon Scullion}
\affiliation{Department of Mathematics, Physics and Electrical Engineering, Northumbria University, Newcastle upon Tyne, NE1 8ST}

\author{James A. McLaughlin}
\affiliation{Department of Mathematics, Physics and Electrical Engineering, Northumbria University, Newcastle upon Tyne, NE1 8ST}

\author{Gavin Ramsay}
\affiliation{Armagh Observatory and Planetarium, College Hill, Armagh, BT61 9DG} 

\author{J. Gerard Doyle}
\affiliation{Armagh Observatory and Planetarium, College Hill, Armagh, BT61 9DG}

\begin{abstract}
We present a detailed analysis of a confined filament eruption and jet associated with a C1.5 class solar flare. Multi-wavelength observations from GONG and SDO reveal the filament forming over several days following the emergence and then partial cancellation of a minority polarity spot within a decaying bipolar active region. The emergence is also associated with the formation of a 3D null point separatrix that surrounds the minority polarity. The filament eruption occurs concurrent with brightenings adjacent to and below the filament, suggestive of breakout and flare reconnection, respectively. The erupting filament material becomes partially transferred into a strong outflow jet ($\sim 60$\,km\,s$^{-1}$) along coronal loops, becoming guided back towards the surface. Utilising high resolution H$\alpha$ observations from SST/CRISP, we construct velocity maps of the outflows demonstrating their highly structured but broadly helical nature. We contrast the observations with a 3D MHD simulation of a breakout jet in a closed-field background and find close qualitative agreement. We conclude that the suggested model provides an intuitive mechanism for transferring twist/helicity in confined filament eruptions, thus validating the applicability of the breakout model not only to jets and coronal mass ejections but also to confined eruptions and flares.

\end{abstract}

\keywords{Solar chromosphere - Solar flares - Solar filament eruptions - Magnetohydrodynamical simulations}

\section{Introduction} \label{sec:intro}

Solar flares are a sudden increase in radiation caused by energy conversion resulting from a rapid reconfiguration of the coronal magnetic field. These events are powerful bursts of radiation with energy outputs sometimes exceeding $10^{32}$ erg and can be observed across the entire electromagnetic spectrum \citep[and references therein]{fletcher2011observational}. Having been studied for over 150 years, the underlying physical mechanisms leading to energy conversion during solar flares remains a focus of investigation. 

Magnetic energy release in solar flares can manifest itself in a number of different observables, notably flare ribbons and post-flare arcades, but also in filament eruptions \citep[eg.][]{rust2001, sterling2005, 2013AdSpR..51.1967S}, coronal mass ejections (CMEs) \citep[eg.][]{moore2001,2005JGRA..11011103E, Karpen2012} and blow-out jets \citep[eg.][]{Moore2010, moore2013, young2014}. It is clear the pre-flare magnetic field topology dictates the manifestation of any number (or all) of these phenomena in a solar flare. Here we focus on the role of a filament eruption in a confined solar flare.  

Filaments (or prominences when observed at the limb) are long-lived, stable features which are present in the solar atmosphere and appear as long, thin, dark structures when viewed against the solar disk \citep[and references therein]{engvold2015description}. However, filaments within active regions tend to be shorter in length, lower in height and have shorter timescales than those present in the quiet Sun  \citep{parenti2014solar}. They consist of relatively cool, dense plasma suspended against gravity by the magnetic field in the corona. Both quiet Sun and active region filaments form along a polarity inversion line (PIL) in photospheric magnetic fields \citep{parenti2014solar, chen2017physics}. They exist in force balance, with the outward magnetic pressure of the filament channel balanced by the downward tension of the strapping field above. Filament eruption follows from the catastrophic loss of this force balance {via resistive processes, e.g. breakout reconnection \citep{1998ApJ...502L.181A,Antiochos1999} and tether-cutting \citep{2001ApJ...552..833M}, and/or via an ideal instability \citep[][and references therein]{Chen2011}, e.g. the kink \citep{2005ApJ...630L..97T} and torus \citep{2006PhRvL..96y5002K} instabilities.}

Confined flares are flare events where the solar atmosphere is bound to the surface and there is no eruption of plasma out into space. Confined flares are often associated with the failed eruption of a filament \citep[e.g.][]{Ji2003}. Simulations and observations have revealed that there are two main scenarios for confined filament eruptions. A pre-existing flux rope in a bipolar active region becomes ideally unstable (usually to the kink instability), but the overlying field is too strong or has a low decay index and halts the eruption \citep[e.g.][]{Torok2005,Hassanin2016,Amari2018,Liu2018}. In some cases the decay index may be high enough, but the development of the instability destroys the coherence of the flux rope before it can erupt \citep{Zhou2019}. The kink instability is associated with the conversion of twist to the dimensionless quantity of writhe, i.e. the measurement of the helical deformation of the flux rope about its axis. Often there is a clear development of writhe in these events \citep[e.g.][]{Ji2003,Torok2005}. Alternatively, the filament forms in a multi-polar topology and the erupting material is redirected along nearby coronal loops by reconnection of the erupting structure, \citep[e.g.][]{Devore2008,sun2013hot,Joshi2014,Reeves2015,masson2017,Yang2018}. These events are associated with multiple flare ribbons, and in particular several have been observed with circular ribbons indicative of a coronal null point topology \citep[e.g.][]{sun2013hot,masson2017}. The reconnection and redirection of the erupting material is also sometimes associated with a jet-like surge of plasma \citep[e.g.][]{Yang2018}.  

Multi-polar confined filament eruptions and their associated redirected plasma flows are locally similar in nature to coronal jets generated by the eruption of so-called ``mini-filaments" -- small-scale filaments typically of length $10$ to $30$\,Mm \citep[e.g.][]{Panesar2016}. Coronal jets are a solar phenomenon with a constant presence throughout the solar cycle and have been observed since the launch of Yohkoh in X-ray emission \citep{shibata1993observations, shimojo1996statistical}. They are commonly found in coronal holes (and also active regions) and posses a collimated, beam like structure originating from coronal bright points. Recent observations have revealed that the majority of coronal jets are generated by mini-filament (or sigmoid) eruptions \citep[e.g.][]{2015Natur.523..437S,Kumar2019}. Typically these jets begin with a brightening at the base followed by rapid helical plasma outflows guided by the surrounding magnetic field. They are smaller than typical flares or CMEs with energies in the range of approximately 10$^{26}$ - 10$^{27}$ erg \citep{pucci2013reconnection}. Overall, jet properties include lengths, velocities and lifetimes which are in the range of 1.5 $\times$ 10$^{5}$ km (large side), 100 - 400 km/s and 100 - 16,000 seconds, respectively.

\citet{wyper2017,wyper2018} developed a three-dimensional simulation model for mini-filament jets in coronal holes, building upon concepts introduced in previous jet/CME simulations and observations \citep[e.g.][]{Shibata1986,Antiochos1999,Lynch2008,Pariat2009,Archontis2013,Moreno-Insertis2013,2015Natur.523..437S}. In their model, surface motions are used to form a filament channel along a section of a quasi-circular PIL beneath a coronal null point. In an analogous manner to how breakout CMEs occur \citep{1998ApJ...502L.181A,Antiochos1999}, breakout reconnection at the null point allows the filament channel to rise, inducing tether cutting reconnection \citep{moore2001} that forms a flux rope if one is not already present. When the flux rope reaches the breakout current layer it is explosively reconnected on to the ambient open field, launching non-linear Alfv\'{e}n waves and driving a helical jet as the twist within the opened section of flux rope propagates away. 

In this study, we are interested in the mechanisms for confined filament eruptions in multi-polar topologies and the links to solar flares and jets. We utilise multi-wavelength observations of a confined filament eruption from the Swedish Solar Telescope \citep{scharmer2003sst}, Solar Dynamics Observatory \citep{Pesnell2012sdo}, Big Bear Solar Observatory and Teide Observatory. These observations provide a unique perspective of the area surrounding the filament eruption and also the inferred magnetic field topology prior to the flare. To aid in our interpretation we refer to a modification of the \citet{wyper2017,wyper2018} jet model, where a filament channel eruption launches a jet confined along coronal loops. The details of the observations are given in \S 2 and \S 3 to \S 5 describe the filament eruption and jet. The observations are contrasted with the MHD simulation in \S 6, where we find excellent qualitative agreement. Finally, \S 7 summarises our interpretation of the observations, whilst in \S 8 we discuss the broader ramifications of our work and present our conclusions. 

\begin{figure}
    \centering
    \includegraphics[width = 0.47\textwidth]{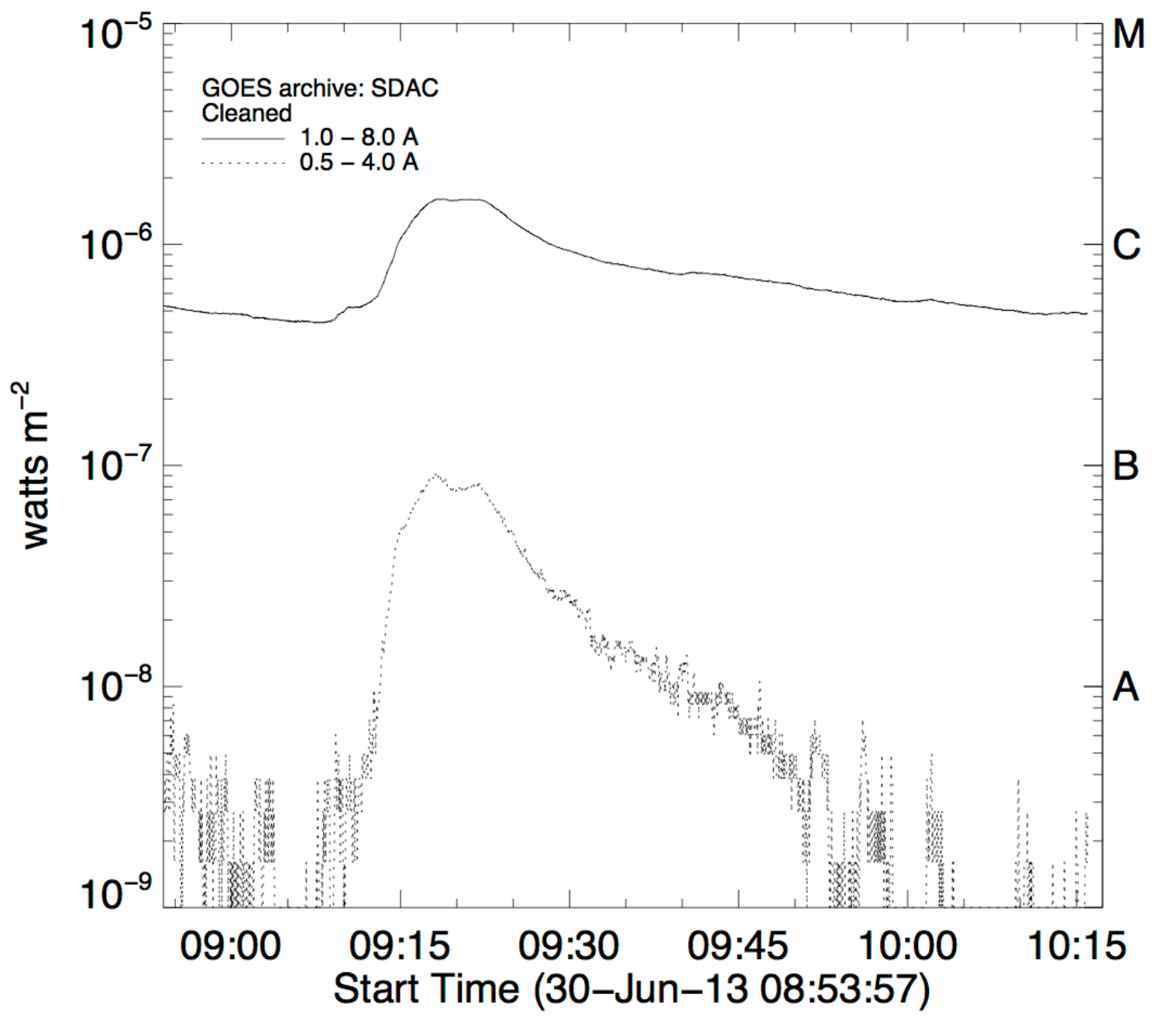}
    \caption{The GOES lightcurve for the time of the SST/CRISP observations which show this flare event as a C1.5 class flare in the 1.0 - 8.0\AA\ channel.}
    \label{goes_light}
\end{figure}

\section{Observations}
\subsection{Ground Based}

The Swedish Solar Telescope (SST) observed a filament eruption associated with a C1.5 class solar flare on the 30th June 2013 in AR 11778 close to the disk centre. The observations were made using the CRisp Imaging SpectroPolarimeter (CRISP) \citep{scharmer2008crisp}, a spectropolarimeter based on a dual Fabry-P{\' e}rot interferometer design which operates in the red beam from 510 - 860nm. It has three cameras, two of which are narrowband of 0.3 - 0.9nm and one wideband. The CRISP spectral scans are centred on the H$\alpha$ 6562.8\AA\ absorption line of the chromosphere. The field of view (FOV) of these observations is approximately 60\arcsec $\times$ 60\arcsec with an image scale of 0.0592\arcsec per pixel. The observations consist of a series of images scanning the H$\alpha$ spectral line in the range of $\pm$1.38 \AA\ with 50 milli-Angstrom equidistant steps resulting in 33 spectral line positions scanned. Overall, the active region was observed for approximately one hour at a temporal resolution of 7.27 seconds with the eruption and flaring occurring within the first 5 minutes. 

The CRISP FOV is corrected for solar tilt and the bright points in the wideband images are cross-correlated with those in SDO Atmospheric Imaging Assembly (AIA) \citep{Lemen2012aia} 1700\AA\ for co-alignment, achieving a sub-AIA pixel accuracy in the CRISP pointing and establishing a heliocentric coordinate system for CRISP. Sub-AIA pixel alignment of the CRISP pointing is achieved as a result of a cross-correlation of the most intense CRISP pixel within the AIA pixel space of 10 coincident bright points (initially identified by eye within a GUI). Then the CRISP pixel space, within the AIA bright point, is explored for each of the 10 bright points in order to maximise the correlation and a correction to the pointing information of CRISP is established. As a result, the CRISP observations are centred on $(x,y) = (323.36\arcsec, -287.91\arcsec)$ with a roll angle of 62.04\degree. Each pixel contains the 33-point spectral scan of H$\alpha$ and this makes up the spectral data cube for investigation using the CRisp SPectral EXplorer (CRISPEX: \cite{vissers2012flocculent}). The standard procedure for the reduction of CRISP is given by \cite{de2015crispred}, and includes a correction for differential stretching. Post-processing was applied to the data sets using the image restoration technique Multi-Object Multi-Frame Blind Deconvolution (MOMFBD), as outlined by \cite{van2005solar}.

Full-Disk H$\alpha$ (FDHA) images were acquired from the Global Oscillations Network Group \citep[GONG:][]{harvey1996global, harvey2011full} hosted by the National Solar Observatory. In particular FDHA images were collected from the Big Bear Solar Observatory (BBSO) and Teide Observatory. FDHA images are taken with a cadence of 1 minute and each 2k $\times$ 2k image has a spatial sampling of 1\arcsec. These context observations are used to identify and monitor the filament emergence from its first appearance until it eventually erupts.

\begin{figure*}[ht!]
    \centering
    \includegraphics[width = 1.0\textwidth]{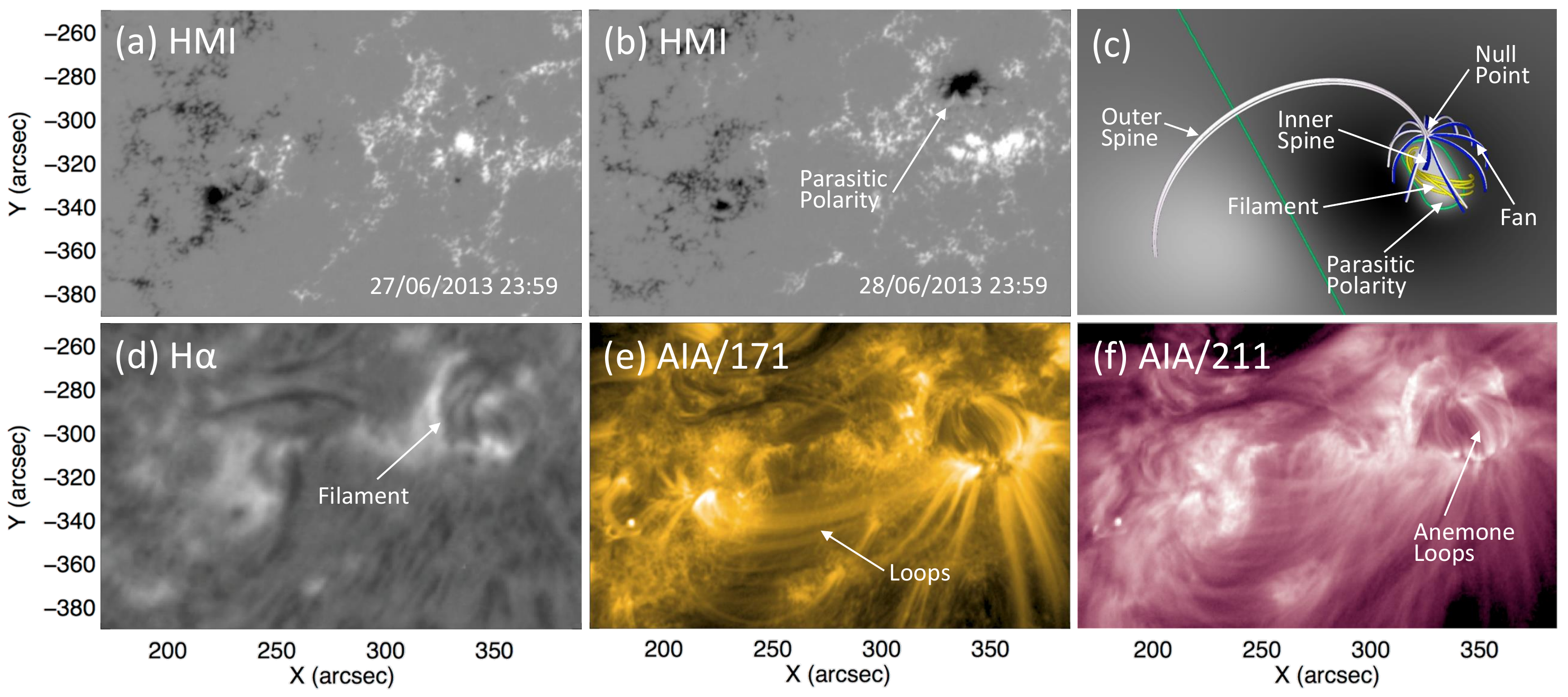}
    \caption{Here we show multi-wavelength observations of the active region where (a) and (b) represent the HMI magnetograms before and after the parasitic polarity appears, respectively. These images are taken 24 hours apart on the 27/06/2013 23:59UT and 28/06/2013 23:59UT. In panel (c) we show the pre-flare magnetic field of the simulation where the green lines denote the PILs. Surface shading shows the normal component of the magnetic field scaled between $\pm 240$\,G. Panels (d) - (f) show the active region after the parasitic polarity emerges at 23:59UT on 28/06/2013 in H$\alpha$, 171\AA\ and 211\AA. The FOV for each of the HMI, AIA and H$\alpha$ images is $220\arcsec \times 140\arcsec$ and is centred on $(x,y) = (280\arcsec, -320\arcsec).$
    \newline An animation is available that shows the sequences in panels (b), (d) and (f). The first 17 seconds of the video is of HMI magnetograms of panel (b) from 27-Jun-2013 15:59:10.7 to 1-July-2013 23:59:10.8. From 18 to 66 seconds the video shows the H$\alpha$ sequence of panel (d). This sequence begins at 28-June-2013 00:01:10 and ends at 30-June-2013 09:43:31. The last 16 seconds of the video is the AIA 211\AA sequence of panel (f). It begins on 27-June-2013 23:30:11.62 and ends exactly one day later. The total real time duration of the video is 83 seconds.}
    \label{pre_flare_structure}
\end{figure*}

\subsection{Space Based}
The GOES \citep{bornmann1996goes} soft X-Ray lightcurve of the C1.5 solar flare is presented in Figure \ref{goes_light}, showing the flare beginning at 09:11UT with a peak at 09:18UT. For analysis of the flare ribbons in this flare event in particular, including Hard X-Ray (HXR) and high energy signatures, refer to \cite{druett2017beam}.  

We utilised data from AIA as well as the Helioseismic and Magnetic Imager (HMI) \citep{Scherrer2012hmi} on board the Solar Dynamics Observatory (SDO). AIA consists of four 20cm dual-channel telescopes which provide multiple simultaneous full-disk filtergram images (image scale of 0.6\arcsec{} in AIA) of emission lines of the corona and transition region. With its 12s temporal resolution and FOV of 41 arcmin, AIA observes in ten different wavelength channels. In this work, we utilise observations across multiple AIA channels, including 131\AA\ ($1.9 \times 10^{5}$ - $2.5 \times 10^{7}$ K), 171\AA ($2 \times 10^{5}$ - $2.5 \times 10^{6}$ K), 211\AA\ (63,000 - $6.3 \times 10^{6}$ K) and 304\AA (40, 000 - $2 \times 10^{6}$ K) where the temperatures correspond to the passbands of each AIA channel according to the AIA response functions. Additionally, the peak temperature of 131\AA\, 171\AA\, 211\AA\, and 304\AA\ are $6.3 \times 10^{5}$ K, $7.9 \times 10^{5}$ K, $1.7 \times 10^{6}$ K and 80,000 K respectively, covering the chromosphere, transition region, corona and flaring regions within the solar atmosphere. This allows us to view the flaring active region, thereby, providing a larger FOV (in comparison with CRISP) and broader context of the overlying magnetic topology and subsequent evolution of the filament eruption prior to the flare. Every 45 seconds HMI provides 1 arcsecond resolution full-disk magnetic flux images. HMI magnetic flux images provide us with a clear interpretation of the magnetic topology of the active region photosphere containing the filament. Data reduction was carried out using {\small SSWIDL} {\tt aia\_prep} for SDO instruments and for GONG we acquired preprocessed data from the online data archive\footnote{\url{http://halpha.nso.edu/}}.

\begin{figure*}
	\centering
	\includegraphics[width = 0.98\textwidth]{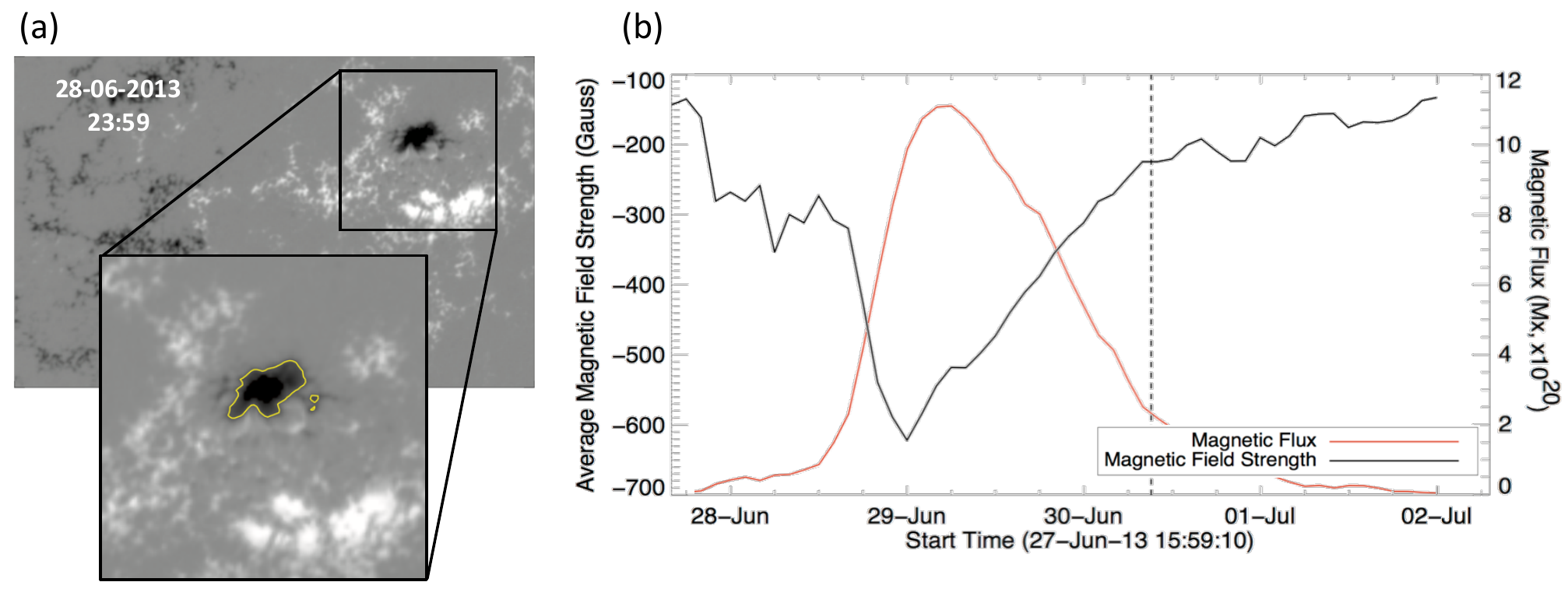}
    \caption{Panel (a) is a zoomed in view of the parasitic polarity detailing the contoured region in yellow as the area used to derive the magnetic field strength and magnetic flux. In (b) we show the average magnetic field strength (black line) of the contoured area of the parasitic polarity and corresponding magnetic flux (red line) with respect to time. The black dashed line represents the flare onset time of 09:11UT.}
    \label{mag_flux}
\end{figure*}

\vspace{15mm}
\section{Pre-eruption Magnetic Configuration}
In order to determine the origin of the flaring event and filament eruption we use HMI magnetograms from SDO together with coronal loop observations from AIA to infer the topology of the active region magnetic field. Two days prior to the flare (27th June 2013) the active region exhibited a simple bipolar photospheric magnetic field footprint. Throughout the 28th June 2013, a patch of negative field (hereafter referred to as the {\it parasitic polarity}) emerges into the positive field region, creating the embedded bipole surface field associated with a coronal null point \citep[e.g.][]{Antiochos1990,Masson2009,sun2013hot,Kumar2018}. Between the 29th and 30th June 2013, the parasitic polarity appears to weaken and fragment as it evolves into the positive field.  

Figure \ref{pre_flare_structure} shows the pre-flare bipolar magnetic field structure of the active region before, (a), and after, (b), the emergence of the parasitic polarity region. Note the negative polarity within panel (a) is a result of transient flux emergence and does not become a part of the parasitic polarity. Panels (d) to (f) show H$\alpha$ and EUV images after the emergence. In the H$\alpha$ movie associated with image (d) we can see a ring filament appearing in the upper right hand corner which corresponds to the location of the PIL surrounding the parasitic polarity in the HMI magnetogram (b). The filament forms and develops over 24 hours and this is simultaneous with the beginning of the flux emergence in HMI (see movie in supplementary material). The EUV images in panels (e) and (f) show that new connections have been formed between the negative parasitic polarity and the surrounding positive polarity in the classic anemone shape associated with a coronal null point \citep{shibata1994gigantic}, see the AIA 211\AA\ movie for more details. From comparing the large-scale coronal loops, the filament position and magnetogram we infer the ring filament has formed beneath the spine-fan topology of a coronal null. An outer spine is therefore expected to follow the large-scale loops and connect to the surface in the negative magnetic flux concentrations on the left hand side of the magnetogram. For an illustrative comparison, in Figure \ref{pre_flare_structure}(c) we show the pre-flare magnetic field structure of the 3D MHD simulation which contains the basic constituent features of the observations described. Note that the surface field polarity is reversed with respect to the observation.  

Figure \ref{mag_flux} describes the temporal evolution of magnetic flux and the average magnetic field strength of the parasitic polarity, spanning before and after the flare. To construct the time profiles in panel (b), we identify the parasitic polarity field concentrations within the enclosed box of panel (a) using intensity contours at the level of -100 Gauss; this is denoted by the yellow contour in (a). Panel (b) shows that the parasitic polarity first appears on 27th June 2013 at 15:59 UT and it grows in size and strength for approximately 1.5 days reaching a peak intensity on 29th June 2013 at 00:00 UT (see movie). 

After this the parasitic polarity begins to fragment and disperse covering a larger area ondisk and by 2nd July 2013 it has disappeared. Overall, it had a total lifetime of 3 - 4 days, with the flare under study occurring approximately 33 hours after the peak in magnetic flux. During the cancelling of magnetic flux there were brightenings in the AIA hot channels at the H$\alpha$ filament location, which may be indicative of small scale magnetic reconnection events.

\begin{figure}
    \centering
    \includegraphics[width = 0.47\textwidth]{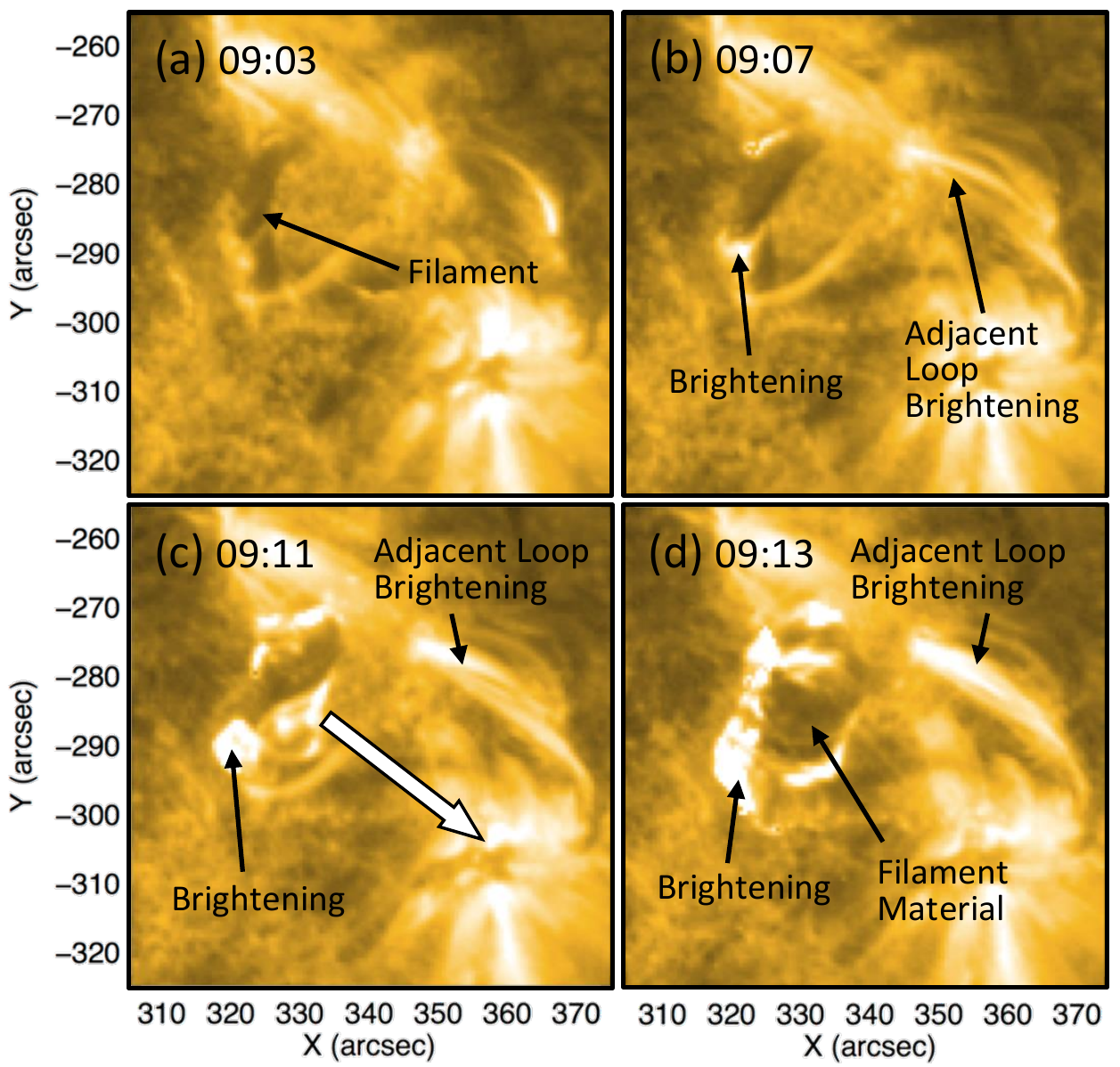}
    \caption{Observations in AIA 171\AA\ of the active region before and during the filament eruption detailing the brightenings adjacent to the filament as a consequence of magnetic reconnection. The white arrow represents the direction of the erupting filament material. The FOV is $70\arcsec \times 70\arcsec$ and is centred on $(x,y) = (340\arcsec,-289\arcsec)$, it is the same FOV as the zoom in box in Figure \ref{mag_flux}(a).
    An animation of the figure is available. The video begins at 30-Jun-2013 09:00:11.34 and ends on the same day 09:19:47.34. The real time movie duration is 20 seconds.}
    \label{filament_loops}
\end{figure}

\begin{figure*}[ht!]
    \centering
    \includegraphics[width = 0.97\textwidth]{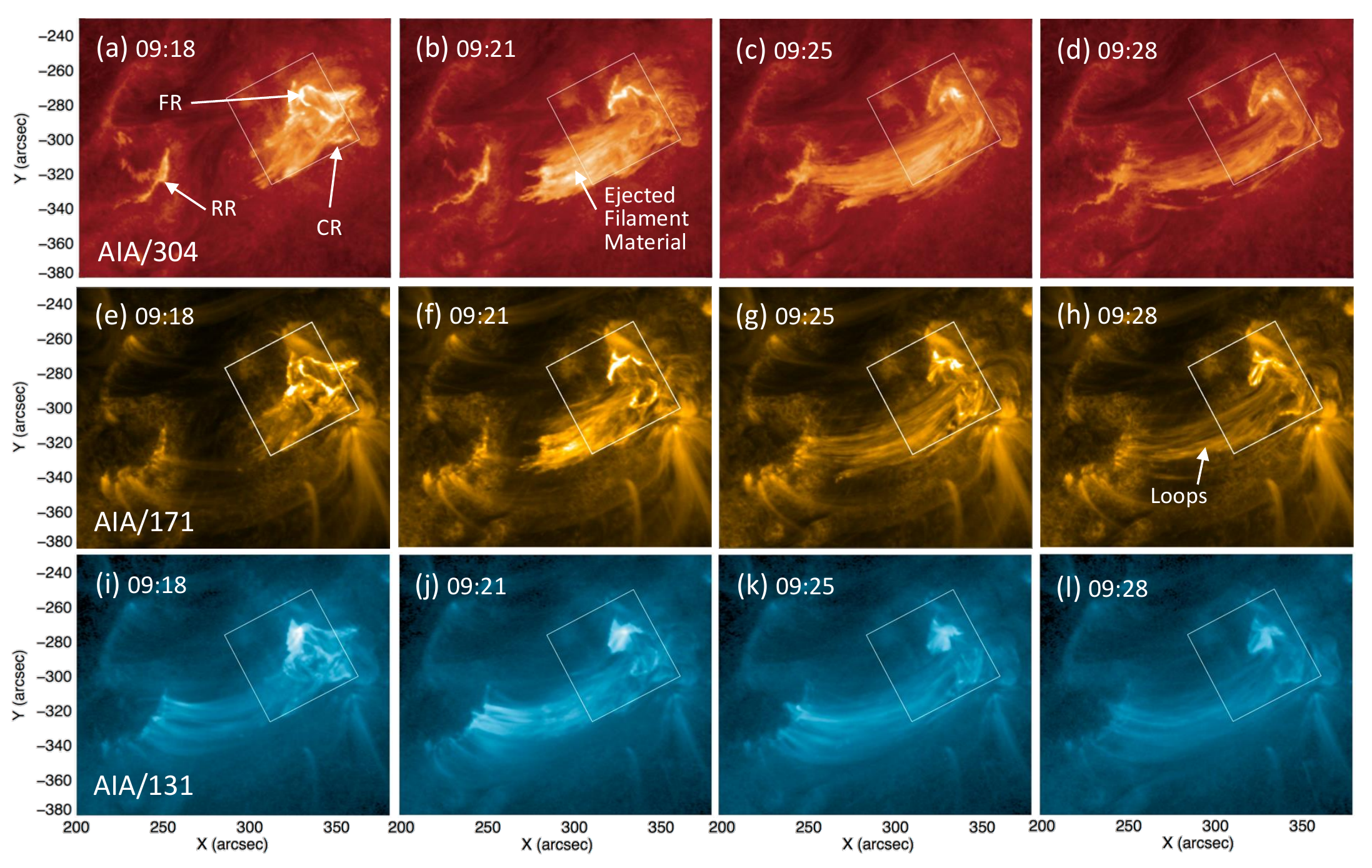}
    \caption{A sequence of AIA images in 304\AA\, 171\AA\ and 131\AA\ spanning a total of 10 minutes after the filament eruption has started. The FOV is $180\arcsec \times 150\arcsec$ and is centred on $(x,y) = (290\arcsec, -305\arcsec)$ and the white box represents the CRISP FOV. Abbreviations are as follows: Circular Ribbon = CR, Remote Ribbon = RR, Flare Ribbon = FR.
    An animation of the figure is available. The video begins at 30-Jun-2013 09:00:11.34 and ends on the same day 09:19:47.34. The real time movie duration is 20 seconds.}
    \label{sdo_transfer}
\end{figure*}

\section{The Filament Eruption}
Figure \ref{filament_loops} is a four image sequence from (a) - (d) spanning 10 minutes before and during the eruption  with respect to 171\AA. Equivalent image sequences in Figure \ref{filament_loops_appendix} are presented in panels (a) - (d) with respect to 304\AA\ and panels (e) - (h) with respect to the hot coronal \lq\lq{flaring}\rq\rq{} line 131\AA{}. For context, the FOV is shown by the zoomed image of the magnetogram in Figure \ref{mag_flux}(a). In panel (a) we observe the clear structure of the filament lying along the left section of the quasi-circular PIL. Four minutes later, as shown in panel (b), the filament has started to erupt. At this time bright loops appear to the right of the rising filament which increase in brightness over time as the filament erupts, panels (c) and (d). Further brightenings are also observed immediately adjacent to the rising filament at these times in 171\AA{} and in 131\AA. As will be discussed further in \S\ref{sim} these brightenings are signatures of breakout and flare reconnection, respectively. Throughout this phase of the eruption the filament material is accelerated  moving from the north east to the south west of the region as we see it, this is clearly seen in the associated movie. 

In Figure \ref{sdo_transfer}, the AIA observations in 304\AA\, 171\AA\ and 131\AA\ detail the next phase of the eruption. Panels (a) - (d) demonstrate that soon after beginning to erupt the filament material is transferred to the extended (overlying) active region coronal loops, where it then propagates eastward in the form of a large-scale outflow (jet) back to the surface, making the eruption completely confined within the active region, see movies. Such a transfer is only possible if the field lines supporting the erupting filament material have been reconnected through the null point. Figure \ref{sdo_transfer}(e) - (l) shows that the jet is also multi-thermal, i.e. also containing a heated plasma component, consistent with this picture. The multi-thermal jet appears to flow within or beneath a set of substantially hotter overlying post-flare loops that already exist at the time of the formation of the jet, i.e. comparing panels (a) and (i).{/bf This is a result of the filament material being presumably denser that the hotter (outer) corona. Therefore, the filament was physically above these hot loops (seen in AIA/131) otherwise the loops would not be visible as their emissions would become absorbed by the filament.}

\begin{figure*}[ht!]
	\centering
	\includegraphics[width = 0.9\textwidth]{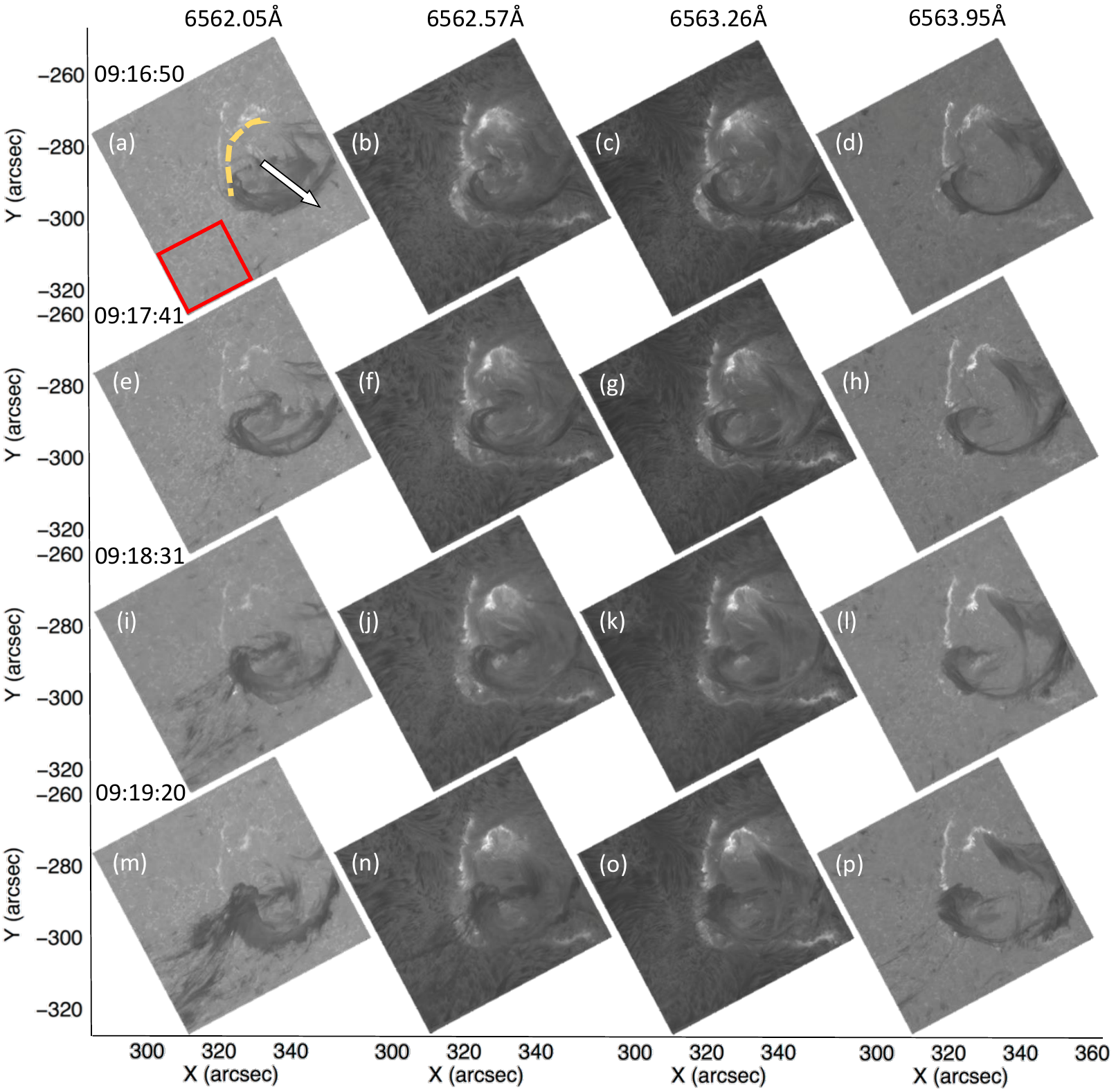}
    \caption{This mosaic of CRISP H$\alpha$ spectral images from (a) - (p) shows the observed filament eruption in varying wavelengths at different times. Each column represent a different wavelength position within the H$\alpha$ line profile. The rows then represent the time stamps which span a total of 2.5 minutes and show the evolving untwisting filament feature. The red box in panel (a) represents the quiet Sun region used to calculate the rest wavelength profile and the dashed yellow line is the filament location prior to erupting. Additionally, the white arrow represents the direction of the erupting filament material. The FOV for each image is $60\arcsec \times 60\arcsec$ and is centred on $(x,y) = (323.6\arcsec, -287.91\arcsec)$. 
    An animation of this figure is available. The video runs from 0-June-2013 09:15:54 to 10:53:08 of the same day. The real time duration of the movie is 17 seconds.}
    \label{sst_crisp}
\end{figure*}

Further evidence for reconnection through the null is provided by the flare ribbons. The brightest are the two parallel ribbons formed by flare reconnection beneath the erupting filament. However, also present is a remote flare ribbon and circular flare ribbon. All three ribbons (FR: Flare Ribbon; CR: Circular Ribbon; RR: Remote Ribbon) are shown in Figure \ref{sdo_transfer}(a). The remote and circular ribbons are the expected signatures of energy deposition in the chromosphere from non-thermal particles accelerated near the null point that escape along the outer spine and fan plane, respectively \citep[e.g.][]{Masson2009}. \cite{druett2017beam} studied the H$\alpha$ response of the southern section of the circular ribbon in this flare event and obtained excellent agreement with a 1D beam electron model.

SST/CRISP captured the crucial moments where the erupting material was transferred to the overlying loops and the jet was launched in excellent detail. Figure \ref{sst_crisp} shows the CRISP spectral image sequence of the filament eruption and flare ribbons, at four times in 50~s intervals, i.e. from 09:16:50 UT to 09:19:20 UT (rows) and at four wavelength positions (columns), sampling the H$\alpha$ spectral profile. In panels (a) - (d), at the beginning of the sequence, the filament eruption is well underway. At this time the erupting filament material has formed an arch-like shape, having already erupted towards the south west (bottom right). The two flare ribbons are also visible near the continuum in the H$\alpha$ wings of panels (a) and (d) which run parallel to the original location of the filament. The legs of the erupting structure appear to be connected to the surface near the ends of the parallel flare ribbons, as one would expect for a typical filament eruption. Also visible near the line core of panels (b) and (c) is the southward section of the circular ribbon which appears to connect to the base of the left-most flare ribbon at this time. The subsequent panels then show that over the next two and half minutes, filament material from around the southern leg of the erupting structure begins to be transferred to the overlying coronal loops, propagating away to the south east. This is most easily seen in the blue wing, e.g. (a), (e), (i) and (m). Accompanying this transfer of filament material is the development of a strong clockwise rotation of the filament structure, seen most easily in the accompanying movie and discussed further below. The above along with the simulation results (discussed below) further supports the conclusion that the flux rope supporting the erupting filament material has been reconnected on to the overlying loops near its apex, transferring plasma from its southward leg. This will be discussed in detail in Section \ref{sim}. 
\begin{figure*}
	\centering
	\includegraphics[width = 0.9\textwidth]{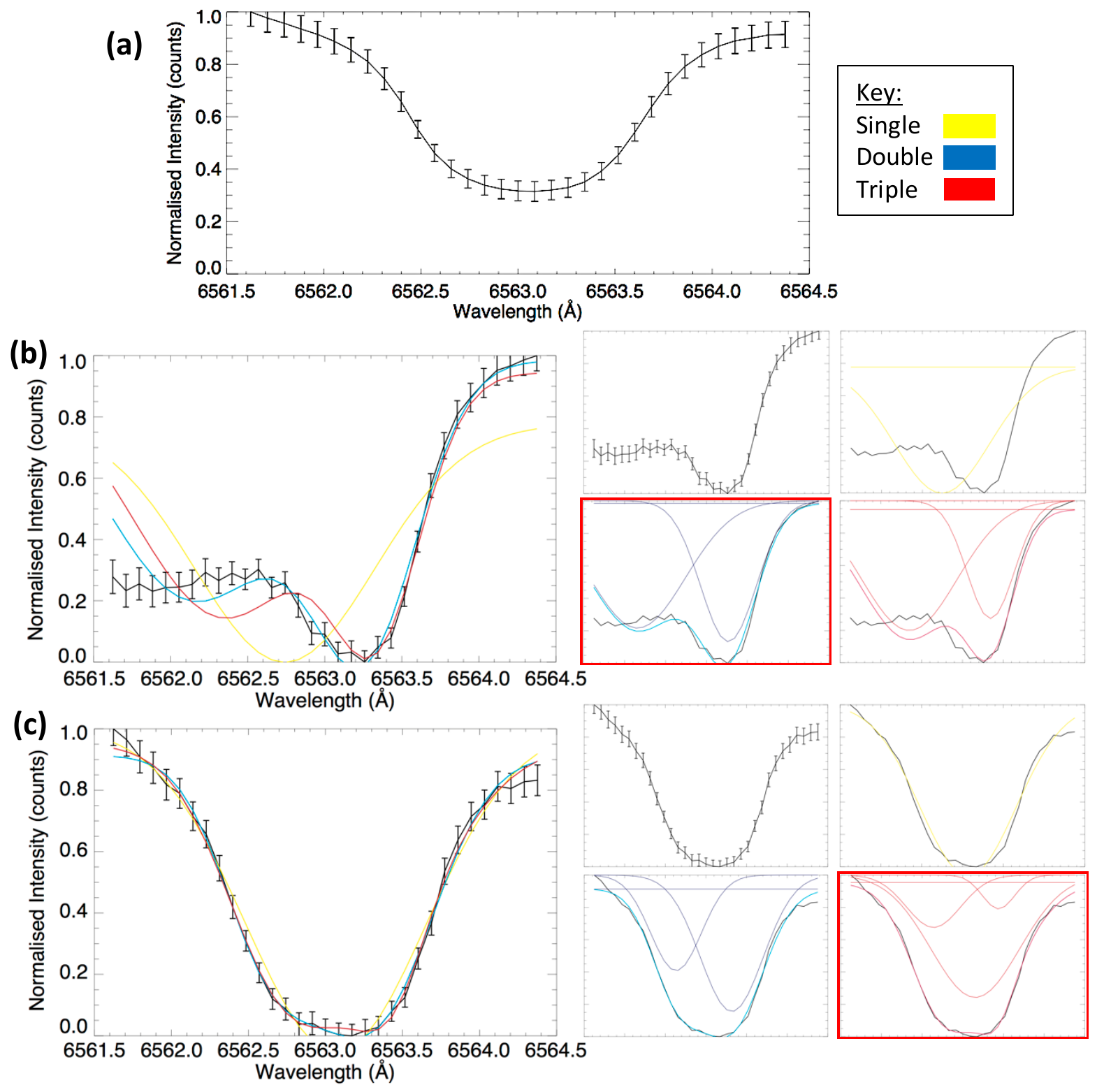}
   
    \caption{Gaussian fits to the H$\alpha$ absorption profiles are presented for a number of cases in panel rows (a) - (c). Panel (a) presents the normalised, rest H$\alpha$ line profile along with the associated errors on each wavelength point. This normalised profile is made up from the averaging of thousands of quiet sun profiles from the last frames in the SST data when the filament has erupted and is not within the FOV. Panel (b) presents spectral line profile fits for a pixel sampling the moving/outflowing filament which exhibits a highly blue-shifted wing component. As a result of this, the double Gaussian model, highlighted within the red-boxed sub-panel provides the best fit to the data. The individual components of each Gaussian model are also presented in the sub panels. Panel (c) presents the line profile fit for a quiet Sun pixel location showing a relatively unshifted H$\alpha$ profile. The triple Gaussian model fits the line best here and is highlighted within the red-boxed sub-panel. }
     \label{Gaussian_fit}
\end{figure*}

\begin{figure*}
	\centering
	\includegraphics[width = 0.97\textwidth]{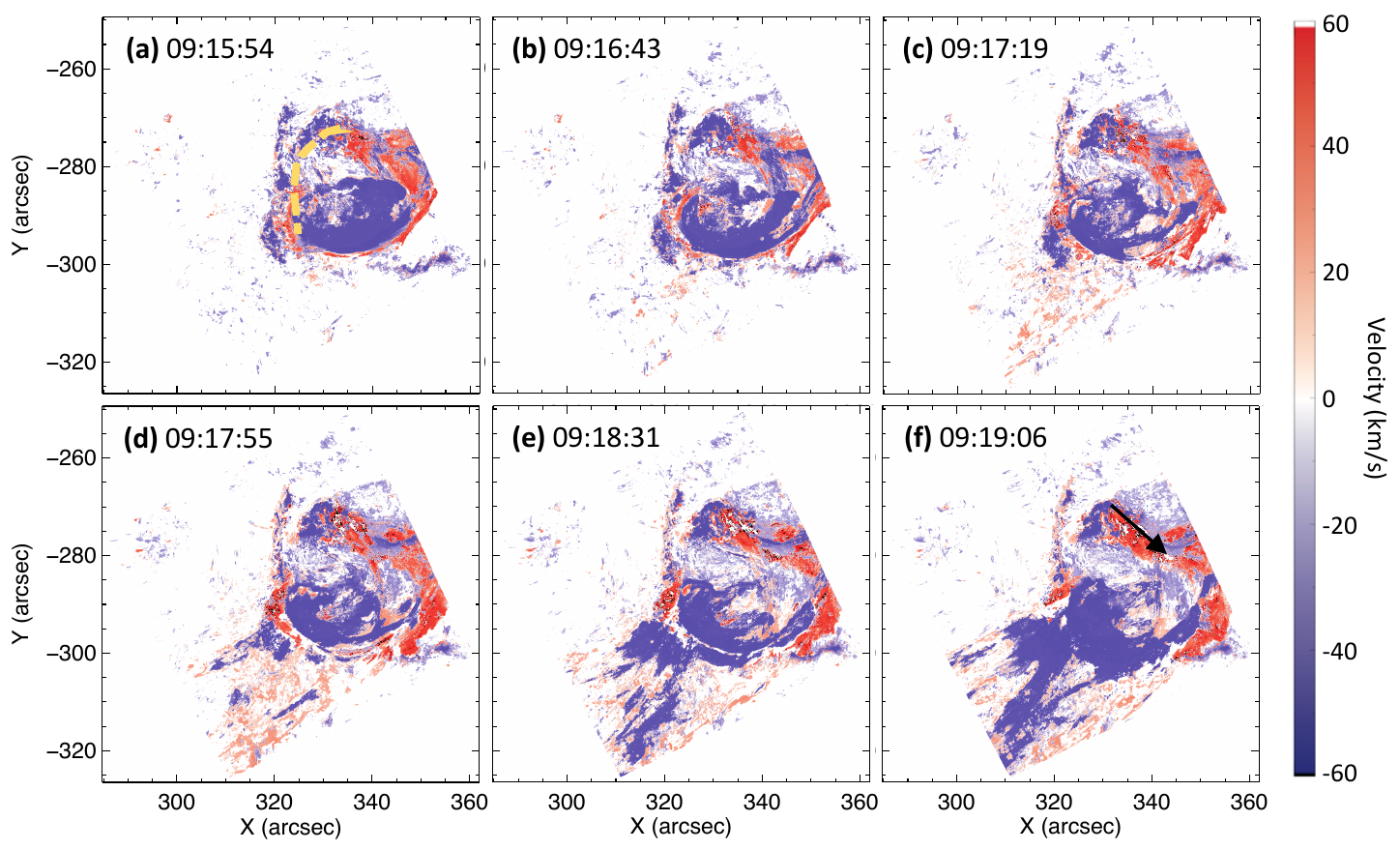}
    \caption{Velocity maps constructed as an amalgamation of the extremely blue-shifted and red-shifted components within the filament eruption. The dashed yellow line represents the filament location prior to erupting and the black arrow the flows which move downwards to the solar surface.
    An animation of this figure is available. The video begins at 0-June-2013 09:15:54 and ends on the same day at 09:19:06. The real time movie duration is 10 seconds.}
    \label{vel_maps}
\end{figure*}

\section{Jet Kinematics from H$\alpha$ profiles}
In order to understand more about the plasma outflow and its development of rotation we carried out a line fitting analysis on the CRISP H$\alpha$ observations. Due to the complex spectral line deformations within various spatial pixels throughout the eruption, several multi-Gaussian fit functions were applied in order to identify which combination of Gaussian functions could achieve the overall best spectral line fit, following a reduced $\chi^{2}$  minimisation test. Every pixel was fitted with a single, double and triple Gaussian and the $\chi^{2}$ statistic was minimised to achieve the best fit to the line profile. For the double and triple Gaussian fits constraints were placed on the centroid wavelength of each Gaussian. This enabled the Gaussian components to fit the various features of the profile, especially when it was highly blue or red shifted. The constraints for the centroids of the double Gaussian are 6561.62 - 6563.1\AA\ and 6563.1 - 6564.37\AA and the triple Gaussian 6561.62 - 6562.82\AA\, 6562.82 - 6563.25\AA\ and 6563.25 - 6564.37\AA. In addition, the background level was set as a standard polynomial fit with zero degrees and was not constrained. The $\chi^{2}$ obtained could then be used to select the most appropriate fit function on a pixel-by-pixel basis and for all spectral scans in time. This process was completed for all pixels in the first 30 time frames of the observation, consisting of the full duration of the filament material transfer.

In order to simplify this process a block fitting routine was implemented, available within {\small SSWIDL}, called \texttt{cfit\_block}. In addition, all the profiles were normalised to each of their maximum y-value, this makes up the background/zero level. Once the iterative fitting process is completed the resulting output is a data structure consisting of the centroid wavelength position, amplitude and FWHM for each Gaussian fit component, for each time frame, from which we can investigate the evolving spectral profiles in greater detail. 

Figure \ref{Gaussian_fit} shows example fits for a pixel within the filament eruption, (b), and a quiet Sun pixel away from the event, (c). The corresponding components of each of the fit functions are also shown, within 4 sub-panels to the right of (b) and (c), for completeness. By applying the reduced $\chi^2$ minimisation method and identifying the fitting combination that maximises the number of zero-line crossings in the goodness-of-fit residuals, we can iterate through all pixels at all times. This process of selection of best fit functional form results in the assignment of a key value from 1 - 3 for each pixel, in order to create a map of the preferred fit as being either single (1), double (2) or triple (3) Gaussian. Further details on the distribution of functional fits for the FOV and more examples of line fits to complex absorption profiles are given in Appendix \ref{app:maps}.

An important parameter obtained from the fitting routine is the centroid wavelengths of each Gaussian component. These wavelengths can be used to calculate the corresponding Doppler velocities of the line profiles which can represent plasma upflows and downflows. To calculate the Doppler shift from the Gaussian fit components a rest wavelength was obtained from the averaged quiet Sun profiles (see Figure \ref{Gaussian_fit}a) summed over a section of pixels in the SST FOV away from the flaring and eruption regions (see Figure \ref{sst_crisp}(a)). Using the rest wavelength of 6563.06\AA\, determined from the rest wavelength profile obtained from the quiet Sun profiles, the Doppler velocity of each pixel at each of the 30 time frames can be computed to produce velocity maps as the erupting filamentary material is transferred into the confined jet. With regards to the triple Gaussian best fit pixels, we can construct Doppler velocity maps of the plasma flows in the blue wing, core and red wing. Further details of how these velocity maps are constructed are given in Appendix \ref{app:maps}. We are primarily concerned with the motions of the erupting filament material we focus on the highly blue-shifted and red-shifted components within the velocity maps, largely arising from double Gaussian best fits. 

\begin{figure*}
	\centering
	\includegraphics[width = 1.0\textwidth]{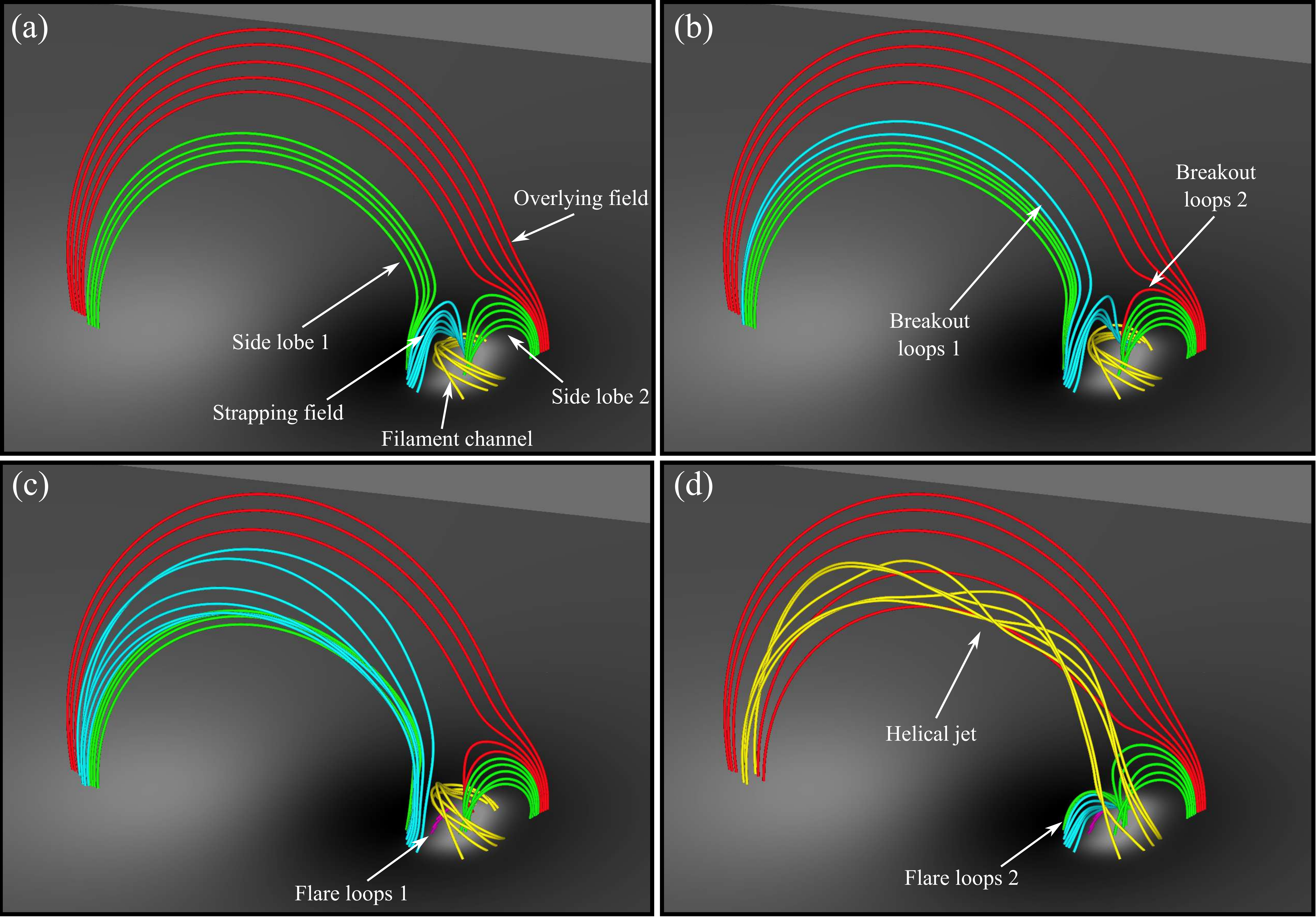}
    \caption{Field lines showing the eruption of the filament in the MHD model. Yellow: the filament channel. Cyan: overlying strapping field. Green: side lobe regions. Red: overlying background field. (a) $t = $12 min 55 s, (b) $t = $14 min 35 s, (c) $t = $16 min 15 s, (d) $t = $17 min 5 s.}
    \label{fls}
\end{figure*}

In Figure \ref{vel_maps} each panel represents a different time step beginning at 09:15:54 UT and ending at 09:19:06 UT. Each panel image represents an amalgamation of the red and blue velocity maps which correspond to the red and blue wing components revealing the locations of the largest outflows. In Figure \ref{vel_maps}(a), the inferred shape of the erupting filament is shown as a dashed yellow line. The top portion of the arch shaped erupting filament is strongly blue-shifted (reaching at least 60 km/s), whilst plasma in the legs to the left and top right is red shifted ($\approx$ +45 km/s). The red shifts show that filament plasma in the legs is moving downward towards the solar surface, whilst the plasma near the top of the arch is being ejected upwards, see movie for more detail. Similar plasma motions are routinely observed in large-scale filament eruptions, \citep[e.g.][ and references therein]{pant2018twisting}. Based on its similarity to large-scale eruptions in this phase, we conjecture that the downflows are predominantly due to mass draining along the legs as they become angled upwards, however other mechanisms such as driving from pressure gradients or magnetic tension can not be discounted. Note that due to the wavelength window available for making these maps, the inferred velocities are effectively limited to $\pm$60 km/s. Therefore, the true line of sight velocity in the strongly blue shifted regions could be much higher.

Over the next few minutes, i.e. panels (b) - (d), the filament material develops fine structure and becomes transferred on to the longer active region loops (flowing towards the south-west in each panel). The transferred material remains strongly blue-shifted and begins to rotate clockwise as it propagates away to the bottom left (the rotation is more evident in the online movie). Although more structured, the red shifts of the right-most leg continue to generally show downflows throughout this time. Downflows near the foot point of the other leg are also visible next to the strongly blue-shifted material in the jet. We therefore find that downward as well as upward motions of the filament material occur during the filament eruption and further confirm our conjecture that reconnection of the erupting flux rope near its southern leg is responsible for launching the filament plasma into the jet. 

\newpage
\section{3D MHD Simulation}
\label{sim}
\subsection{Setup}
To explore our conjectures further we conducted a 3D MHD simulation with the Adaptively Refined Magnetohydrodynamics \citep[ARMS code: ][]{Devore2008} for qualitative comparison with the filament eruption and jet in the observed event. The key details of the simulation setup are described below. For further specifics see Appendix \ref{app:sim}.

The simulation was initialised with a uniform background plasma and a potential magnetic field containing a large-scale bipole with a small-scale embedded parasitic polarity. The resulting field has a similar 3D magnetic null point topology to the one inferred from the observations. The system was then energised using surface motions which formed a small-scale filament channel beneath the null point in a similar position to the observed filament. This method of creating the filament channel is simply a numerically convenient way of introducing the free magnetic energy where we want it and is not meant to reproduce how this particular filament was formed. The surface driving is then halted once the filament channel has formed, and the system allowed to evolve without external forcing from this point onward. Field lines showing the filament channel and the spine-fan topology of the null point in the simulation just before eruption are shown in Figure \ref{pre_flare_structure}(c).

\subsection{Eruption evolution}
The simulated filament channel eruption proceeds in the same manner as the coronal hole jet simulations reported in \citet{wyper2017,wyper2018}. Figure \ref{fls}(a) shows the filament channel (yellow field lines) prior to eruption. Four other field line regions are also shown: cyan -- overlying strapping field; green -- side lobe regions (1 and 2) and red -- overlying background field. The null point resides where the four regions meet each other, with the closed outer spine following the path between the overlying (red) and side lobe 1 (green) field lines. 

The increasing magnetic pressure within the filament channel expands the overlying strapping field upwards, quasi-statically balancing the outward magnetic pressure with magnetic tension. However, this expansion also stresses the null point so that a current sheet forms there (the breakout sheet). Reconnection within the sheet then slowly transfers the strapping field to the side lobe regions, reducing the downward magnetic tension force on the filament channel and allowing it to rise, Figure \ref{fls}(b). This leads to a faster rise, which in turn leads to faster breakout reconnection (the breakout feedback mechanism). Although the simulation does not include dense chromospheric plasma and the effects of gravity, we would still expect a similar qualitative evolution in a magnetically dominated low-$\beta$ plasma. However, the timing and speed of the filament rise might be expected to alter slightly.

The reconnected strapping field forms new coronal loops in the two side lobe regions, labelled breakout loops 1 and 2 in Figure \ref{fls}(b). Loops 1 form nearby to the original position of the outer spine, whereas loops 2 form beneath the domed fan plane separatrix. The rising filament channel also stretches the strapping field, creating a flare current layer beneath it. Tether cutting/slipping flare reconnection within this layer converts the sheared arcade into a twisted flux rope, whilst also forming short flare loops beneath, Figure \ref{fls}(c) (flare loops 1). 

Although not explicitly included in the simulation, one would expect enhanced coronal EUV emission associated with both the flare and breakout loops. As such, the origins of the observed brightenings shown in Figure \ref{filament_loops}(b) during the filament eruption now become clear. The adjacent loop brightening shows new loops formed by the breakout reconnection of the strapping field, i.e. breakout loops 2. Interestingly, at this time there is no clear coronal EUV signature of loop heating associated with breakout loops 1. This maybe because the deposited energy from the breakout reconnection is spread over a much larger volume than that for breakout loops 2, reducing the intensity of emission. The brightening beneath the filament shows bright, heated plasma within the flare loops. Such energy release beneath the erupting filament material implies flare-like reconnection has set in beneath the erupting material. This flare reconnection forms a flux rope in our model (in common with all eruptive flare models). Therefore, we can infer that if a flux rope was not already present prior to eruption, it will be by this point in the evolution.

\begin{figure*}
	\centering
		\includegraphics[width = 0.97\textwidth]{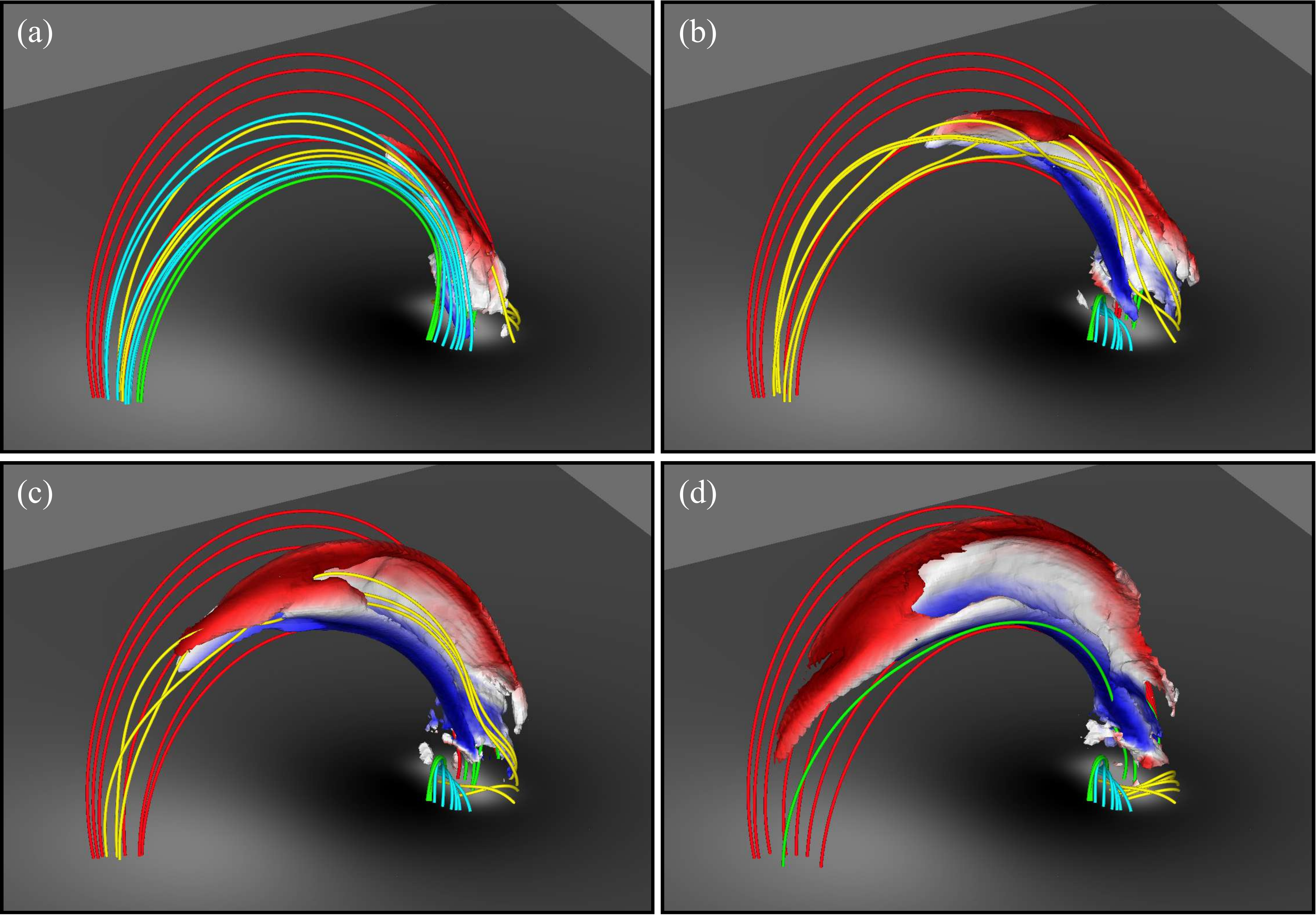}
    \caption{An isosurface of velocity ($|v| = 500$\,km/s) showing the jet. The isosurface is shaded to show the $v_{z}$ value (saturated at $\pm 500$\,km/s), showing the jet rotation. (a): $t = 16$\,min $40$\,s. (b): $t = 17$\,min $5$\,s. (c): $t = 17$\,min $30$\,s. (d): $t = 17$\,min $55$\,s.
    An animation of this figure is available. The video runs from t = 0 min to t = 24 min 10 secs. Its real time duration is 10 seconds.}
    \label{vel_iso}
\end{figure*}

Returning to the simulation, the rise of the flux rope then accelerates as the feedback loop sets in between the flux rope rise and the removal of strapping flux via breakout reconnection. This continues until all of the strapping field is reconnected away, Figure \ref{fls}(c). Such a feedback loop could explain the steadily increasing intensity of the flare and adjacent loop brightenings shown in Figure \ref{filament_loops}(c) - (d). Beyond this time in the simulation the rising flux rope itself begins to reconnect with the overlying field. The end of the flux rope rooted in the background negative polarity is then connected on to distant closing field lines, whereupon the twist begins to propagate along the loops as a non-linear Alfv\'{e}n wave, Figure \ref{fls}(d) (yellow field lines). The end of the flux rope rooted in the minority polarity reforms the filament channel, but now with a reduced shear (not shown). We inferred a similar evolution of the erupting flux rope in the observed event from the filament plasma evolution in the SST/CRISP images (Figs. \ref{sst_crisp} and \ref{vel_maps} and associated movies). That is, in both the simulation and observation a rotating jet is produced when the erupting structure is reconnected, transferring its twist to the overlying field.

One further aspect of the field line evolution that is worthy of mention is that once the erupting flux rope is reconnected, the flare reconnection after this time transfers the flux moved into the side lobe regions in the breakout phase back into the overlying field and strapping flux regions, Figure \ref{fls}(d) (red and cyan field lines). As discussed in detail in \citet{wyper2018}, once the flux rope is reconnected the null in the breakout current layer moves into the flare current layer beneath the erupting structure. Thus, the new flare loops formed after this time (flare loops 2) are through null point reconnection as opposed to tether cutting/slipping reconnection prior to this. Observationally, the signature of this transition should be that one of the parallel flare ribbons becomes part of the circular ribbon, as the two are now formed by energy deposition from the same reconnection region. This is precisely what we observed in the H$\alpha$ (Figure \ref{sst_crisp}) and EUV 304\AA\ (Figure \ref{sdo_transfer}(e)) in our event as the jet is launched.

\subsection{Helical Jet}
Figure \ref{vel_iso} shows an iso-surface of velocity depicting the plasma jet formed by the transfer of twist in the simulation. Qualitatively, the jet is very similar to the observation in that (i) it is helical in nature, (ii) it is guided along the ambient coronal loops back to the surface and (iii) it is formed from a mixture of ambient coronal plasma swept up by the reconnection and plasma from within the filament channel (see movie).

Quantitative comparison of the speeds however reveal that the simulation jet is significantly faster. In the simulated jet the plasma propagates at roughly the local Alfv\'{e}n speed within the loop, reaching speeds of $\approx$ 450 km/s with typical coronal scaling values. This is higher than in the observation, where values of $60$ km/s were recorded, Figure \ref{vel_maps}. However, as noted earlier the wavelength window used for constructing the velocity maps leads to an effective cap of $\pm$60 km/s for the inferred velocities, with the real value expected to be higher. Additionally, the simulation uses a simplified atmosphere with a uniform background plasma and neglects gravity. As such, although the magnetic field structure of the filament channel is formed and evolved in a qualitatively correct manner, it does so in the absence of the denser, cooler filament material seen in the observation. It would not be unreasonable to expect that were such dense plasma be included, the propagation speed along the loop might be reduced by the locally slower Alfv\'{e}n speed and the action of gravity. A more sophisticated simulation would be required to test this claim. Despite this, the close qualitative comparison between the observations and simulation, incorporating a simplified model atmosphere, serves to highlight the pertinence of reconnecting magnetic fields in dictating the overall dynamics of this event and the potential universality of this model in sufficiently describing a variety of similar events within differing atmospheres and on different scales.

\section{Event Summary}
\begin{figure}
	\centering
		\includegraphics[width = 0.47\textwidth]{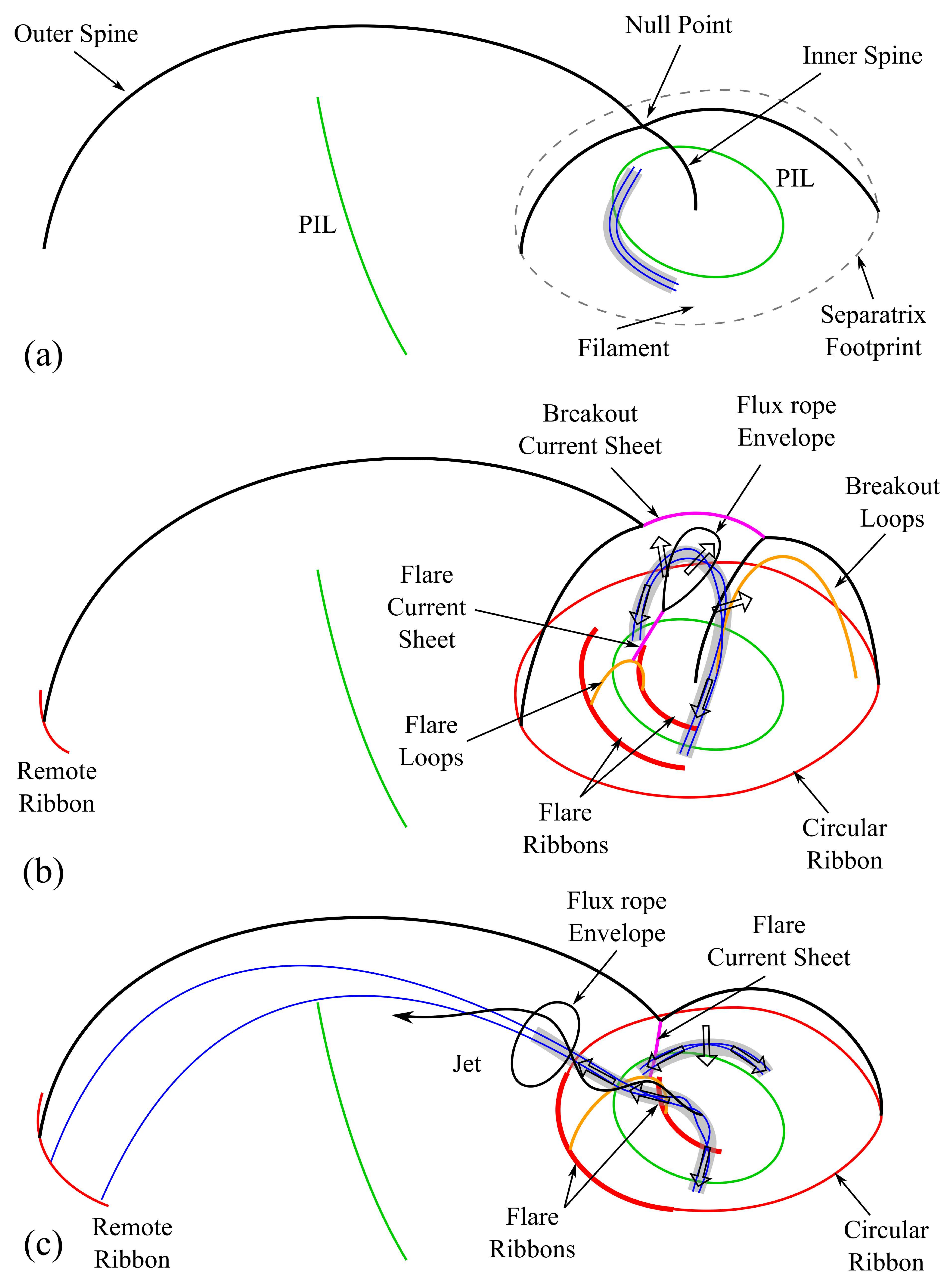}
    \caption{Schematic of the eruption. (a) pre-eruption. (b) during the breakout phase. (c) after the filament is reconnected and the jet is launched.}
    \label{schem}
\end{figure}

In this work we presented a detailed analysis of a confined filament eruption/flare and its associated helical jet. Using observations from SDO/AIA/HMI, GONG and SST/CRISP we studied the formation of the filament and its surrounding magnetic topology, the eruption and the subsequent jet kinematics. In particular, the SST/CRISP observations gave us a detailed view of the transfer of filament material as the jet was launched. Qualitative comparison with a 3D MHD simulation of a closed-field breakout jet further aided our interpretation of the observations. Figure \ref{schem} shows a schematic which summarises our interpretation of the different stages of the confined eruption. 

Figure \ref{schem}(a) shows the configuration just prior to eruption. The filament (dark grey) resides along a section of the quasi-circular PIL (green) beneath the separatrix of the 3D null point. This configuration forms over $\approx 1.5$ days as the parasitic polarity emerges. The evolution of magnetic flux in the parasitic polarity (Figure \ref{mag_flux}) suggest that a combination of flux emergence and cancellation is involved in the formation of the filament.

Figure \ref{schem}(b) shows the filament mid-eruption. The null point above the filament has collapsed into a breakout current sheet (pink). Breakout reconnection removes strapping field from above the filament, forming bright loops adjacent to the erupting material (breakout loops, orange). New loops should also be formed nearby the outer spine, but they are much less intense and not clearly observed in our event. Additionally, the upward stretching of the strapping field forms the flare current sheet beneath the filament material (pink). Reconnection in the flare sheet forms a flux rope if not already present (or adds further polodial flux if one is present initially) and bright flare loops (orange). The flux rope forms an envelope (cross section shown in black) around the erupting filament material which resides in its core. The material near the core is accelerated upwards with the erupting flux rope, whilst the material in the legs falls back to the surface (hollow arrows). Energy deposition from the breakout current layer creates circular and remote ribbons at the base of the fan and outer spine, respectively (thin red lines). Parallel flare ribbons also form at the base of the flare loops that stretch between the two feet of the erupting flux rope (thick red lines).

Once the erupting flux rope reaches the breakout current layer it is reconnected, Figure \ref{schem}(c). This splits the flux rope near its apex, forming a new shorter closed loop along which downflows of filament plasma are observed (e.g. top right arrow, Figure \ref{vel_maps}(f)). The other section of flux rope is now connected to coronal loops with foot points nearby the original foot point of the outer spine. As the twist within the flux rope propagates along the loops, it drives a mixture of cooler, denser filament material and hotter, more tenuous heated plasma along these loops as a helical jet. At this point the flare current layer has reached the separatrix, so that the flare reconnection is actually null point reconnection. One of the parallel flare ribbons now forms a section of the circular flare ribbon, whilst the other outlines field lines near the foot point of the inner spine (thick red lines), e.g. Figure \ref{sdo_transfer}(a).

\section{Discussion \& Conclusions}
The observations presented herein are consistent with the breakout picture for jet generation. However, we can not rule out the possibility that the triggering mechanism differs from breakout and is instead an instability of the flux rope itself. Figure \ref{filament_loops} (and the associated movie) showed that the brightening associated with breakout reconnection occurred simultaneously with the brightening from flare reconnection. This could be consistent with slow, low energy breakout reconnection prior to the eruption that speeds up once flare reconnection is initiated (as in \citet{Karpen2012,wyper2018,Kumar2018} for example). Or equally the breakout reconnection could be reactionary, following from the eruption of the filament driven by an ideal instability (as suggested by e.g. \citet{masson2017} for their event). The partial cancellation of the parasitic polarity flux suggests that a flux rope may have formed in the filament channel prior to the eruption, thus both of the above scenarios are a possibility. A non-linear force-free extrapolation of the pre-eruption magnetic field could potentially help to pin down the pre-eruptive field structure and aid in diagnosing the eruption trigger. However, this is outside the scope of the present work which is focused on understanding the eruption kinematics.

Regardless of the exact trigger, once the eruption is underway it is clear that breakout reconnection is heavily involved in the eruption as demonstrated by the similarities with the simulation and the high resolution CRISP and AIA observations. Similar to jets involving mini-filaments in coronal holes, the strength of the overlying field suppresses ideal expansion of the flux rope once flare reconnection ensues. Without being able to blast the overlying field outwards, breakout reconnection of the strapping field and then ultimately the erupting flux rope provides the only avenue to eject the twist/helicity from the filament channel. In the case of open-field coronal hole jets the twist then propagates away along open field lines, whereas in these confined events it becomes trapped on overlying loops. 

Why then do all confined flares in null topologies not show clear evidence of associated jets? This is likely to do with the relative size of the separatrix surface compared with the surrounding coronal loops. \citet{wyper2016} quantified this with the ratio $L/N$, where $L$ is the distance between the two spine foot points and $N$ is the width of the footprint of the separatrix dome on the solar surface. They found in simulations of jets driven by rotating the parasitic polarity that in configurations where $L/N \approx 1$ minimal jets were produced, whereas for larger ratios the jets became more defined and higher energy. The classic jet producing topology of a parasitic polarity surrounded by open field corresponds to $L \to \infty$ (and therefore $L/N \to \infty$), consistent with this picture. Recently, \citet{masson2017} studied a confined flare where $L/N \approx 1$ finding little evidence of clear outflows, whereas in the event studied by \citet{Yang2018} we estimate that $L/N \approx 2.1$ and a clear rotating jet spire was observed. In our event, based on the pre-flare EUV loops (Figure \ref{pre_flare_structure}(e) and (f)), we estimate that $L/N \approx 2.6$.

Our event and simulation are relatively small compared to some confined events. \citet{Devore2008} studied homologous breakout eruptions that are much larger in scale. However, in a similar manner to our breakout jet simulation they find that a full scale breakout eruption and CME is suppressed when a strong overlying field is present. The erupting filament channel is instead reconnected across the breakout current layer, transferring its shear/helicity to the overlying field. The key differences from the present model is the larger scale of their simulation (allowing greater ideal expansion of the filament channel) and an $L/N$ ratio of 1, giving less coherent jet-like outflows. However, the basic physics is the same in the two models. In this sense, the current simulation bridges the gap between the large-scale confined eruption simulations of \citet{Devore2008} and the open-field jet simulations of \citet{wyper2017,wyper2018}.

Taken together with previous studies of jets \citep[e.g.][]{wyper2017,Kumar2019}, CMEs \citep[e.g.][]{Lynch2008,Karpen2012,Chen2016} and other confined filament channel eruptions (with and without associated jets) \citep[e.g.][]{sun2013hot,Masson2009,Yang2018} our results support the conclusion that all of the above phenomena can be tied together by the shared topology of a filament channel formed beneath the separatrix of a coronal null. In such a configuration, breakout reconnection can and should be expected to be involved in the eruption. The present investigation demonstrates that in the context of confined filament eruptions, the breakout process provides and intuitive mechanism for confining the eruption by redirecting it along overlying field.

\section*{Acknowledgements}
LD acknowledges funding from an STFC studentship (ST/N503927/1). PW is supported by a RAS fellowship. Armagh Observatory and Planetarium is core funded by the Northern Ireland Government through the Dept. for Communities. The authors would like to thank the staff of the SST for their support with the observations. We thank Spiro Antiochos, C. Richard DeVore and Judy Karpen for stimulating discussions and consultation on the numerical simulation. ES and JM acknowledge STFC via grant number ST/L006243/1 and for IDL support. Finally, we would like to the thank the anonymous referee for their comments and suggestions which helped to imporve the paper. The Swedish 1-m Solar Telescope is operated on the island of La Palma by the Institute for Solar Physics at Stolkholm University in the Spanish Observatorio del Roque los Muchachos of the Instituto de Astrofisica de Canarias. SDO Data supplied courtesy of the SDO/HMI and SDO/AIA consortia. SDO is the first mission to be launched for NASA's Living With a Star (LWS) Program. This work utilizes $H\alpha$ intensity data obtained by the Global Oscillation Network Group (GONG) project, managed by the National Solar Observatory, which is operated by AURA, Inc. under a cooperative agreement with the National Science Foundation.

\bibliographystyle{aasjournal.bst}
\bibliography{solar_flare_paper.bib} 

\appendix
\renewcommand\thefigure{\thesection.\arabic{figure}}
\setcounter{figure}{0}

\section{Additional SDO/AIA Images}
\begin{figure*}[!h]
\centering
\includegraphics[width = 0.97\textwidth]{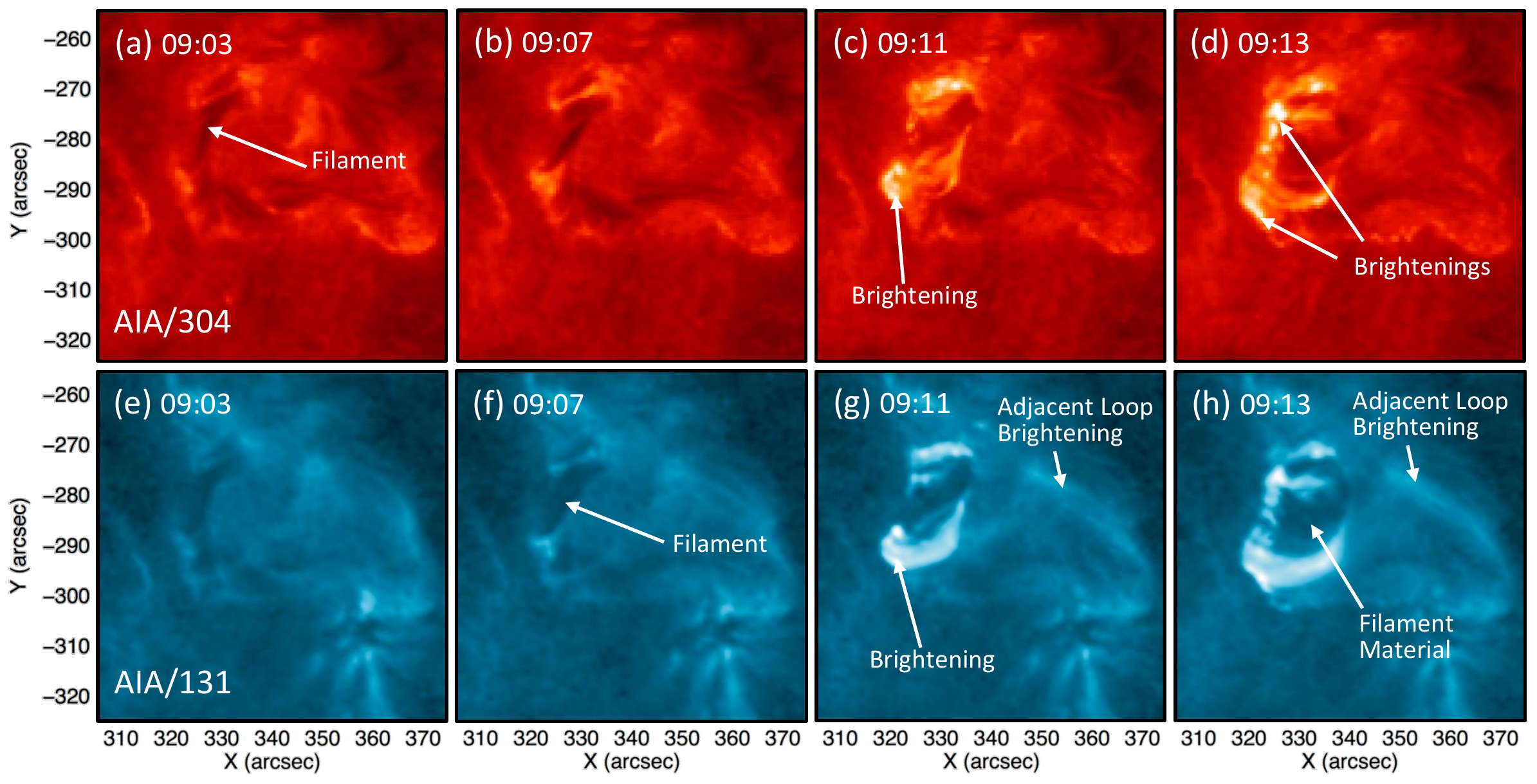}
\caption{Observations in AIA 304\AA\ and 131\AA\ of the active region before and during the filament eruption detailing the brightenings adjacent to the filament as a consequence of magnetic reconnection. The FOV is $70\arcsec \times 70\arcsec$ and is centred on $(x,y) = (340\arcsec,-289\arcsec)$, it is the same FOV as the zoom in box in Figure \ref{mag_flux}(a).}
\label{filament_loops_appendix}
\end{figure*}

In this Figure \ref{filament_loops_appendix} below we provide additional image sequences of the filament eruption from Figure \ref{filament_loops} showing both a chromospheric and a hot flaring line. In panels (a) and (e) we see the clear structure of the filament which rises to erupt in panels (b) and (f). Upon this eruption, brightenings in both 304\AA\ and 131\AA\ are observed in panels (c) and (g) as a result of magnetic reconnection within the region. Finally, in panel (h) we see faint signatures of the brightening adjacent loops which are clearly identified in Figure \ref{filament_loops} in 171\AA.

\newpage
\section{Producing Line-of-sight Velocity Maps}
\label{app:maps}
\begin{figure*}[!h]
	\centering
	\includegraphics[width = 0.9\textwidth]{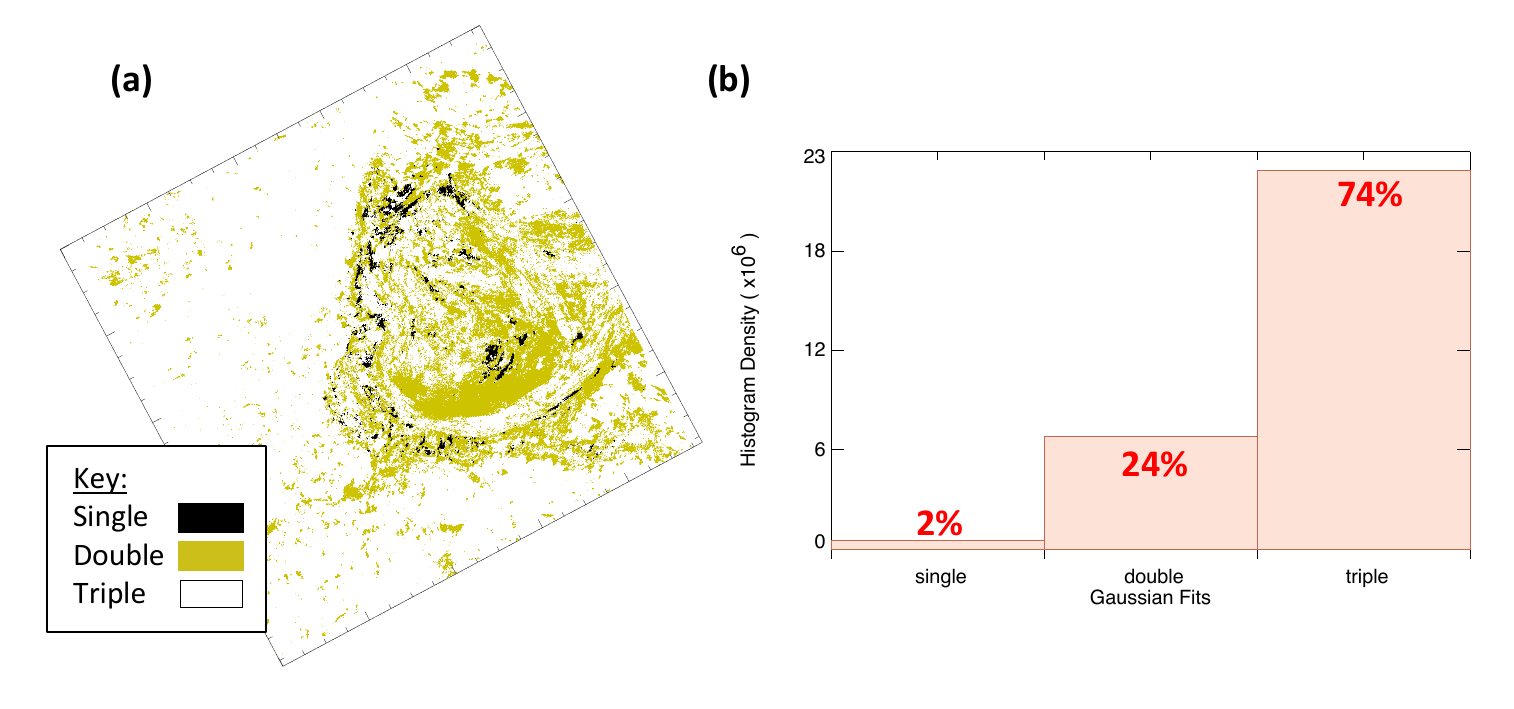}
    \caption{The two figures convey the statistics of the fitting method used. (a) A snapshot of which features within the SST/CRISP FOV prefer particular Gaussian fits. It can be seen that the majority of the filament eruption is fitted with a double Gaussian, whereas the background quiet Sun requires a triple Gaussian. (b) A histogram detailing the percentage of pixels in all time frames in the SST/CRISP FOV which preferred each of the three fits.}
    \label{key_vel}
\end{figure*}

Producing the velocity maps involves a selective process which iterates through all pixels at all times and assigns each pixel a key value from 1 - 3 which, informs whether the pixel is best fitted by a single, double or triple Gaussian, according to the $\chi^2$ minimisation together with maximising the number of residual crossings. This `key', shown in Figure \ref{key_vel}(a), can then be used to construct velocity maps in the blue wing, core and red wing. Where a single Gaussian was preferred the core map would be assigned the velocity component and the blue and red wings would be 0 km/s, as there is no component of large blue or red shift in these profiles. If a triple Gaussian was preferred then each Gaussian component was assigned to the blue, core or red wing. In addition, if the amplitude of the Gaussian in the blue or red wing was less than a background intensity it was removed from the maps. This was designed to eliminate all the random small velocity movements in the background quiet Sun (i.e. spicules etc.) as we are only concerned with the large velocity movements associated with the extended line wings corresponding to blue and red shifted components of the H$\alpha$ component of the jet. Lastly, if a double Gaussian was preferred, and if centroids lie within the FWHM of the H$\alpha$ profile then the velocity is calculated from an average of the two centroids and applied to the core map. Similarly, if one of the centroids lies within the FWHM it is applied to the core map and the remaining Gaussian centroid lies outside the FWHM it is assigned to the red or blue maps depending on its location. 

In Figure \ref{key_vel}(b), the histogram details which Gaussian fits were preferred for the line profiles from all pixels during the first 30 time frames where the untwisting jet occurs. As you can see the triple Gaussian is the most preferred fit to the line profiles with 76\%, double 24\% and single 2\%. From the key image, Figure \ref{key_vel}(a), we can see the triple Gaussian fits are located in the background quiet Sun regions which is to be expected. These profiles are very similar to the rest H$\alpha$ profile and by providing more free parameters  reduces the $\chi^2$. Double Gaussian fits were preferred in the region of the filament where the highly red and blue shifted plasma is located and single fits in the locations of the flare ribbons. This is a result of the fact that the core chromosphere is being emptied during the event at the location of the filament so, the core red and predominantly blue will become subsequently shifted which, then becomes best fitted with two distinct Gaussians that neglect the rest wavelength intensity. Overall, this histogram justifies the significance of the fitting method implemented and the need for multiple fit functions.

\begin{figure*}
	\centering
	\includegraphics[width = 0.9\textwidth]{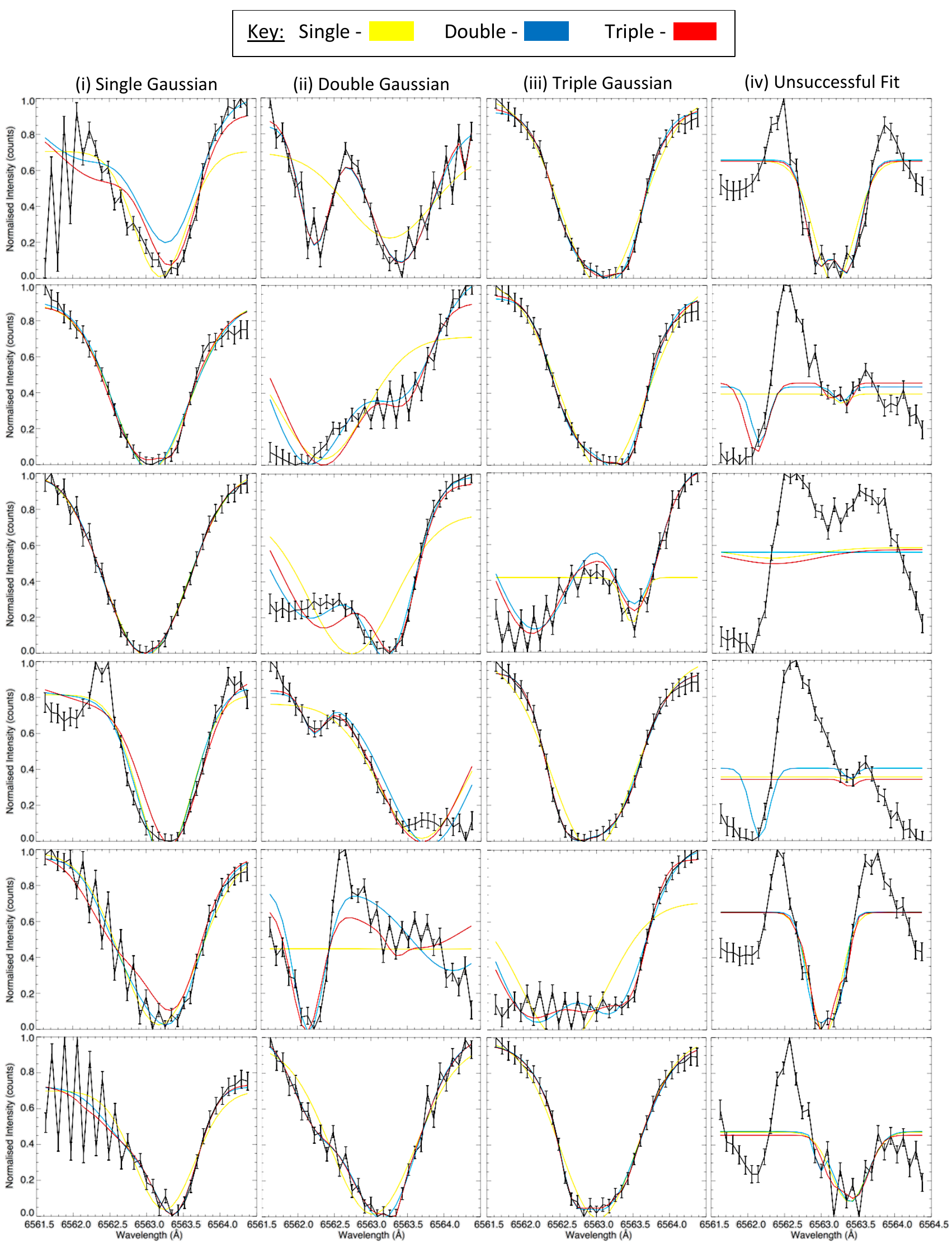}
    \caption{This selection of H$\alpha$ line profiles along with their respected model fits shows the varying degrees of complex profiles and how the fitting method responds. Each column shows profiles which prefer a single, double and triple Gaussian fit with the last column representing unsuccessful fitted profiles. These unsuccessful profiles are a result of the flare ribbons which produce H$\alpha$ in emission and cause issues within the fitting method resulting in none of the Gaussian fits being suitable. However, despite this we are not interested in the fitting of the ribbons as we are concerned with the kinematics of the erupting filament and jet so these profiles are insignificant.}
     \label{pixel_sample}
\end{figure*}

For full disclosure of the profile fitting of the H$\alpha$ absorption line in this study we show a wide variety of single, double, triple and unsuccessful fits in Figure \ref{pixel_sample}. 

\section{MHD Simulation Setup}
\label{app:sim}
The potential field initial condition is constructed using multiple sub-surface magnetic dipoles in the manner of \cite{wyper2016,wyper2018}. The positive polarity of the small bipole forms a minority polarity patch relative to the large-scale background field. This creates a 3D null point spine-fan topology with a distantly closing outer spine, qualitatively similar to the field inferred from our observed event, Figure \ref{pre_flare_structure}. However, it should be noted that the surface magnetic fields are reversed in sign. The null point is situated at $\approx 7$\,Mm above the photosphere, the average width of the domed fan plane is $N = 31.9$\,Mm and the distance between the foot point of the spines is $L = 76.5$\,Mm ($L/N \approx 2.4$). We neglect gravity and assume a uniform background atmosphere ($T = 2.4$\,MK, $\rho = 3.6\times 10^{-15}$\,g/cm$^3$). Gravity is neglected for simplicity as the observed eruption is expected to be dominated by magnetic forces in a low-$\beta$ corona. The qualitative similarity of the simulation and observation show this to be a reasonable assumption after the fact.

Near the apex of the outer spine the plasma-$\beta$ is $\approx 5.4\times 10^{-2}$, whilst $\beta \approx 6.3\times 10^{-4}$ in the centre of the minority polarity. The simulation grid is similar to the one used in \citet{wyper2016}, and contains a maximally resolved region around the seperatrix dome and outer spine. The boundaries are closed and line-tied, with the side boundaries placed far enough away to have minimal effect on the jet dynamics. The ideal, adiabatic MHD equations are solved on the adaptive mesh. Reconnection is therefore through numerical diffusion. The form of MHD equations used is given in \citet{wyper2016}. 

The system is energised using surface motions confined within the parasitic polarity. The flow follows contours of the magnetic field component normal to the surface ($B_{x}$) and therefore do not change the normal surface flux distribution. Details of the flow can be found in \citet{wyper2018}. The fastest surface flows occur along the straight section of PIL around the minority polarity nearest the PIL of the background field (where the highest gradient in $B_{x}$ occurs). This preferentially creates a filament channel there (as in \citet{wyper2018}). The driving is ramped up over a short period ($2$\,min $5$\,s), held constant (until $t = 13$\,min $45$\,s) and then ramped down to zero (at $t = 15$\,min $50$\,s) prior to the jet. The peak driving speed is sub-sonic ($\approx 20\%$ of the sound speed) and highly sub-Alfv\'{e}nic ($\approx 0.5\%$ of the local Alfv\'{e}n speed) so the energy in the filament channel is built up quasi-statically. 

\end{document}